\documentclass[apj]{emulateapj}

\shorttitle{Mizuki}
\shortauthors{Tanaka}

\begin{document}

\title{
  Photometric Redshift with Bayesian Priors on
  Physical Properties of Galaxies
%  ver. \today
}


\author{
  Masayuki Tanaka\altaffilmark{1}
}
\altaffiltext{1}{National Astronomical Observatory of Japan, Osawa 2-21-1, Mitaka, Tokyo 181-8588, Japan}


%\altaffiltext{1}{Visiting Astronomer, Cerro Tololo Inter-American Observatory.
%CTIO is operated by AURA, Inc.\ under contract to the National Science

%-------------------------------------------------------------------
\begin{abstract}
We present a proof-of-concept analysis of photometric redshifts with Bayesian
priors on physical properties of galaxies.  This concept is particularly suited
for upcoming/on-going large imaging surveys, in which only several broad-band
filters are available and it is hard to break some of the degeneracies
in the multi-color space.  We construct model templates of galaxies using
a stellar population synthesis code and apply Bayesian priors on physical
properties such as stellar mass and star formation rate.
These priors are a function of redshift and they effectively evolve
the templates with time in an observationally motivated way.
We demonstrate that the priors help reduce the degeneracy and deliver
significantly improved photometric redshifts.  Furthermore, we show that
a template error function, which corrects for systematic flux errors in
the model templates as a function of rest-frame wavelength,  delivers
further improvements.   One great advantage of our technique is that
we simultaneously measure redshifts and physical properties of
galaxies in a fully self-consistent manner, unlike the two-step measurements
with different templates often performed in the literature.
One may rightly worry that the physical priors bias the inferred galaxy
properties, but we show that the bias is smaller than systematic uncertainties
inherent in physical properties inferred from the SED fitting and
hence is not a major issue.  We will extensively test
and tune the priors in the on-going Hyper Suprime-Cam survey and
will make the code publicly available in the future.
\end{abstract}


%-------------------------------------------------------------------
%-------------------------------------------------------------------
\section{Introduction}

Galaxies form at high density peaks of statistical density
fluctuations in the Universe.  They are thus statistical objects in nature,
and a large survey of the sky is the key observation to
measure their physical properties.  This has motivated recent surveys such as
Sloan Digital Sky Survey \citep{york00},
UKIRT Deep Infrared Sky Survey \citep{lawrence07},
Canada-France-Hawaii Telescope Legacy Survey, and many others.
Following them, there are a number of on-going/planned
surveys, including the Hyper Suprime-Cam Survey (Miyazaki et al.
in prep.), VST and VISTA public surveys, the Dark Energy Survey
\citep{des05}, Euclid\footnote{http://sci.esa.int/euclid/},
WFIRST \citep{spergel13}, and Large Synoptic Survey Telescope \citep{ivezic08}. 
These surveys are going to observe a large fraction of the sky
down to unprecedented depths in order to address some of
the outstanding astrophysical questions today, such as
the nature of dark energy, with a superb statistical accuracy.

Most of these large surveys are imaging surveys.
The information we can
obtain directly from imaging data is unfortunately very limited;
positions, apparent fluxes, apparent sizes, and shapes in a given set of filters,
and their time variability if multi-epoch data are available.
In order to study the nature of distant objects, we need to translate
these apparent quantities into physical quantities
such as luminosities and physical sizes.  Distance 
information is required for this job.  Also, weak-lensing cosmology,
which is one of the major science goals of the large surveys,
requires redshifts (or at least a mean redshift) of source galaxies.
A precise redshift can be measured from a spectroscopic observation.
It is, however, practically not possible to measure precise redshifts
even for 1\% of objects from these surveys with any of
the existing/near-future facilities. Furthermore, most of the objects
will be fainter than the spectroscopic sensitivity limits.
There are currently two techniques to overcome
these problems: photometric redshift and clustering redshift.

Photometric redshift is a technique to infer redshifts of objects
from photometry in a set of filters, first demonstrated by \citet{baum62}.
Thanks to the progress in numerical techniques, it now has two branches:
spectral energy distribution (SED) fitting and machine-learning.
The first branch is the traditional technique and it relies on a priori
knowledge of SEDs of galaxies.  Extensive discussions on the technique can be
found in \citet{walcher11}.  Our new code presented here falls in this
category and we will elaborate on the (dis)advantages of the SED fitting later.
A number of photo-$z$ codes have been published and the most popular ones
would include {\sc BPZ} \citep{benitez00}, {\sc ZEBRA} \citep{feldmann06},
{\sc LePhare} \citep{arnouts99,ilbert06}, and {\sc EAZY} \citep{brammer08}.
The second branch is purely numerical (i.e., no physics) and the idea is to relate
observables such as colors to redshifts in one way or another
using a training set, which has to represent an input photometric sample.
A number of numerical techniques have been applied such as polynomial fitting
\citep{connolly95,hsieh14}, artificial neural network \citep{collister04} and
prediction trees \citep{kind13}.
Extensive comparisons between some of these public SED-based/machine-learning
photo-$z$ codes have been carried out by \citet{hildebrandt10}.

Clustering redshift utilizes the positional information, not flux information,
of objects.  The idea is simple --- cross-correlation between two
galaxy samples yields a signal where they overlap in redshift and
the clustering signal can thus be used as a redshift inference.
Several authors have explored this technique with slightly different
objectives (e.g., \citealt{schneider06,newman08,erben09,mcquinn13,menard13}).
One powerful application of this technique is to use a sample of
spectroscopic redshifts, in which the redshift distribution is precisely known,
as a reference.
If one makes a narrow redshift bin with a spectroscopic sample and
cross-correlate it with a photometric sample, the clustering signal
is proportional to the number of photometric galaxies in that redshift
bin.  By shifting the spectroscopic redshift bin, one can in principle reconstruct
a redshift distribution of the input photometric sample (see \citealt{menard13}
for application of this technique for a few specific cases).
However, there is one uncertainty here; the clustering
signal is also proportional to the bias of the photometric sample,
which is not straightforward to measure a priori \citep{deputter14}.
This can be a serious issue when the photometric sample covers a wide
range of redshift or has multiple peaks in the redshift distribution,
which may often be the case in real analysis.

There are pros and cons in these techniques and one should
choose one that is best suited for his/her science application.
We  focus on the traditional SED fitting technique in
this paper for two reasons: (1) machine-learning techniques require
a representative training sample, which in practice means they
do not go fainter than the spectroscopic limits  and (2) the
clustering technique does not give redshifts of individual galaxies
(it gives $dN/dz$ of the input photometric sample)
\footnote{
  In principle, one can make a dense grid on the color-magnitude space
  and apply the clustering technique to construct $P(z)$ for each grid.
  But, it will require a huge number of spectroscopic redshifts and is
  unlikely a practical method at this point.
  It will not work for rare objects either.
}.
The 1st point is a serious limitation of the machine-learning techniques
because objects that we are concerned about are often fainter
than the spectroscopic limit.  One can of course extrapolate to faint
magnitudes, but an extrapolation always requires
a great care.  With the SED fitting technique, we can 
expect to go fainter than the spectroscopic limit provided that
our understanding of SEDs of galaxies is reasonable.
The 2nd point is also important because, if one
cannot infer redshifts of individual galaxies, one cannot estimate
their physical properties, making galaxy science difficult.

The SED fitting technique has its own problem; it relies
on our a priori knowledge of SEDs of galaxies.
Observed SEDs of galaxies are often used as templates, but
these templates are almost exclusively collected at $z=0$.
The multi-color space at high redshifts may not be fully
covered by those templates because high redshift galaxies
are younger than galaxies at $z=0$.
One can instead use a stellar population synthesis (SPS)
code to remedy the issue with its flexibility to generate
SEDs of a given age.  However,  SPS models do not perfectly
reproduce observed colors of galaxies due to systematics in
the models (e.g., \citealt{maraston09}).  Furthermore, it is
easy to generate models that are physically unreasonable
such as a template with zero SFR with a large amount of extinction.
Such templates increase the degeneracy in the multi-color space,
resulting in poor photo-$z$ accuracies.

In this paper, we choose to use SPS models with 
a novel approach to overcome these problems;
(a) we apply Bayesian priors on physical properties of galaxies to
constrain the parameter space of SPS models within realistic ranges
as a function of redshift and
(b) correct for systematic flux errors using template error functions
(which are an extended version of template error functions often adopted
in the literature).
This technique is still in its early phase of
development and the paper presents a proof-of-concept analysis of
the physical priors and template error functions.

The layout of the paper is as follows.  We give a brief overview of
the paper in Section 2, followed by a description of how
we generate model templates in Section 3.  Section 4 presents
the framework of our technique and describes the physical priors in detail.
We define a template error function in Section 5 and then move on
to demonstrate how these priors and template error functions improve
photo-$z$'s in Section 6.  We compare our code with {\sc bpz} used
for CFHTLenS in Section 7.  Section 8 aims to characterize biases
in inferred physical properties of galaxies introduced by the priors
and template error functions.  Finally, we discuss future directions
in Section 9 and summarize the paper in Section 10.
Unless otherwise stated, we adopt a flat universe with
$\rm H_0=70\ km\ s^{-1}\ Mpc^{-1}$, $\rm\Omega_M=0.27$ and $\Omega_\Lambda=0.73$
\citep{komatsu11}.  
As one of the major
goals of upcoming/on-going surveys is weak-lensing cosmology, it is
important to emphasize that our photo-$z$'s are not dependent on
a specific choice of cosmology.
We simply need to assume a popular cosmology often adopted
in the literature in order to use the observed relationships between
galaxy properties as priors (see Section 4 for details).

%-------------------------------------------------------------------
%-------------------------------------------------------------------
\section{Overview of our photo-$z$ technique and the scope of the paper}

First of all, we give a brief overview of our technique in this section.
Thanks to the enormous amount of work on galaxy populations over
a wide range of redshift, we now have a reasonable
understanding of galaxy properties across cosmic times.
We apply Bayesian priors based on our knowledge of galaxy properties
to SPS models, in which physical properties such as 
stellar mass are known, to constrain the model parameters within
realistic ranges.  We let the priors evolve with redshift
in order to account for evolutionary effects.
A fixed set of templates are often used in photo-$z$ estimates, but
the priors effectively evolve the templates in an observationally motivated manner.

Our approach here is suited for upcoming/on-going large imaging surveys
such as HSC, DES and LSST, which image the sky only in a limited
number of filters in the optical wavelengths.  If one has photometry
in a large number of filters spanning a wide wavelength range with
sufficient S/N, which
may often be the case in deep fields, one does not necessarily have to
apply priors because there is enough information in the data to
constrain the overall spectral shape of objects.  In fact,
\citet{ilbert09} did not apply any priors because they had 30-band
photometry.  However, with the limited photometric information
in the large surveys, it is difficult to break some of the degeneracies
in the multi-color space, resulting in poor photo-$z$ accuracies.
We demonstrate below that the priors developed here are
particularly useful to break (or least weaken) the degeneracies
in such situations and they significantly improve photo-$z$'s.

In addition to the physical priors, we introduce a template error function,
which comes in two terms; systematic flux stretch and flux uncertainty as
a function of rest-frame wavelength.  The first term is to reduce mismatches
between SPS SEDs and observed SEDs and the 2nd term assigns uncertainties
to the model templates to properly weight reliable parts of templates.
The template error function is not a new idea and has been used by
{\sc eazy} \citep{brammer08} and {\sc fast} \citep{kriek09}.
But, they include only the 2nd term.  We introduce the 1st term
in order to
reduce systematic offsets in $|z_{spec}-z_{phot}|$ and improve the overall
photo-$z$ accuracy as we demonstrate in Section 6.

One of the strengths of our code is that it infers, in addition to
redshifts, physical properties of galaxies in a self-consistent manner.
A popular procedure to infer the physical properties is to first
estimate photo-$z$'s with empirical or PCA templates,
and then change templates to
SPS models to infer physical properties.
SPS models are usually not used to infer photo-$z$'s because they
deliver poor accuracy.  But, as we demonstrate in Section 6, 
the physical priors and template error functions improve SPS-based
photo-$z$'s and
we do not need take this 2-step approach anymore.
We can infer both redshifts and physical properties
simultaneously using the same set of templates.  This is important
because redshift and SED shapes are partially degenerate and thus
they have to be measured in a self-consistent manner.
With our technique, we can properly propagate uncertainties in redshift
into uncertainties in physical properties.

Our code is called {\sc mizuki}.  A preliminary version of the code
was used in \citet{tanaka13a,tanaka13b} and \citet{santos14}.  This paper
gives a full description of the code as it stands today.  It is still
in an early phase of development and one of the aims of the paper is to
identify areas where further work is needed.
We focus on galaxies in this paper, but our goal is
to compute relative probabilities that an object is star, AGN/QSO, or
galaxy with corresponding $P(z)$ for each class\footnote{
$P(z)$ for star is simply $P(z)=P_{star}\delta(z)$,
where $\delta$ is the Dirac delta function.}.
Tests on AGNs and stars are being carried out
and we defer detailed discussions to our future paper.

%-------------------------------------------------------------------
%-------------------------------------------------------------------
\section{Stellar Population Synthesis Templates}

We use spectral templates of galaxies generated with the \citet{bruzual03} code.
To the first order, a specific choice of SPS code is not a major concern because
a template error function described in Section 5 will reduce systematic
differences between the model and observed SEDs of galaxies.
In fact, we have confirmed that an updated version of the code, which incorporates
a revised treatment of thermally pulsating AGB stars, deliver similar photo-z accuracies.
We choose to use the older version because there is a growing body of evidence
for thermally pulsating AGB stars being unimportant in integrated
optical-nearIR spectra of galaxies (e.g., \citealt{conroy10,kriek10,zibetti13}),
although these stars may be hidden by dust and may be an important
contributor to thermal emission in the IR.

One needs to make several assumptions to generate SEDs with an SPS code.
These assumptions include, initial mass function (IMF), attenuation curve,
star formation history, and metallicity distribution.  The first two are
often assumed to be time-invariant.  None of them is known a priori
for a given galaxy, and the assumptions employed
inevitably introduce systematic uncertainties in the physical
parameters such as stellar mass inferred from SED fitting.
We choose to employ popular assumptions often adopted in the literature
so that we can use the published results on galaxy properties as
priors (see the next section for details).
To be specific, we assume the Chabrier IMF \citep{chabrier03},
Calzetti attenuation curve \citep{calzetti00}, exponentially decaying
star formation rates (i.e., $\tau$-model), and solar metallicity.
\citet{bruzual03} models do not include nebular
emission lines and so we add them using the intensity ratios in
\citet{inoue11} with the \citet{calzetti97} differential attenuation law.
The Lyman $\alpha$ escape fraction is assumed to be 10\% at all redshifts.
Thermal dust emission is yet to be included in the models, but it does
not affect results presented in this paper.  Absorption due to
the Lyman alpha forest is applied following \citet{madau95}.

Let us briefly discuss the validity of some of these assumptions.
A realistic spectral model should consider a metallicity distribution,
but we assume that all the stars have solar metallicity as often assumed
in the literature.
\citet{conroy09} showed that broad-band
evolution of multi-metallicity galaxy SEDs is equivalent to that of
single mean metallicity population in the redward of the $V$-band.
We should keep in mind that we suffer from multi-metallicity effects,
in addition to the evolutionary effects, at high redshifts where
we often observe rest-frame UV of galaxies.
The biggest uncertainty in our models is perhaps star formation histories.
It is a fundamentally difficult problem to reconstruct star formation
histories from broad-band photometry due to the fact that a small
amount of recent star formation can dominate the overall SED, hiding
light from old stellar populations \citep{maraston10}.
The only way to solve the problem might be to resolve galaxies into
individual stars, but that is not a practical solution for galaxies
at cosmological distances.  Given this fundamental difficulty,
we prefer to be simple and take the popular assumption of $\tau$-models.

We generate SPS templates for ages between 0.05 and 14 Gyr with
a logarithmic grid of $\sim0.05$~dex.
For $\tau$, in addition to
the single stellar population model (i.e., $\tau=0$) and constant
SFR model (i.e., $\tau=\infty$), we assume
$0.1\ {\rm Gyr}<\tau<11\ {\rm Gyr}$ with a logarithmic grid of $0.2$~dex.
Optical depth in the $V$-band goes between 0 and 2 with a step of 0.1
with an addition of $\tau_V=2.5,\ 3,\ 4,$ and 5 models to cover very
dusty sources.  These templates are then redshifted to 
$0<z<7$ with $\Delta z=0.01$ and are convolved with response functions
of a given instrument to generate a library of synthetic fluxes.
A synthetic broad-band flux is computed as

\begin{equation}
  <f_\nu> = \frac{\int d\nu S_\nu f_\nu / \nu }{\int d\nu S_\nu / \nu},
\end{equation}

\noindent
where $S_\nu$ is the response function of an instrument (which
includes the atmospheric transmission for ground-based facilities),
and $f_\nu$ is the input spectrum.
In total, we have about 2 million templates.

%-------------------------------------------------------------------
%-------------------------------------------------------------------
\section{Physical Priors}

We now describe the physical priors.  We begin with the Bayesian
framework and then move on to describe each of the physical priors.

%---------------------
\subsection{Framework}

We follow the Bayesian framework to constrain the parameter space of
the SPS models.  Let $\vec{m}$ be an array of input photometric information
in a given set of filters.  We aim to infer photometric redshift ($z_{phot}$)
from $\vec{m}$.  Physical properties of an input SPS model can be specified
by

\begin{equation}
  \vec{G} = \vec{G}(Z, \tau_V, \tau, age),
\end{equation}

\noindent
where parameters on the right are 
metallicity, optical depth in the $V$-band (i.e., amount of attenuation),
$e$-folding time scale of star formation rate, and age of the model.
At this point, a model is normalized such that initial SFR is
1 $\rm M_\odot\ yr^{-1}$.  Let us introduce a normalization factor,
$\alpha$, in order to be able to compute stellar mass and SFR, which
scale linearly with the normalization.  Stellar mass and SFR are important
because they can be measured more precisely in observations than
the other parameters and we will extensively use them in what follows.
A posterior probability of finding $z$ and $\vec{G}$ from a given
$\vec{m}$ is

\begin{equation}
  P(z,\vec{G}|\vec{m}) \propto \int d\alpha P(\vec{m}|z,\vec{G},\alpha) P(z, \vec{G}, \alpha).
\end{equation}

The prior on redshift and physical properties of galaxies,
$P(z, \vec{G}, \alpha)$, needs some consideration.
Ideally, we would like to apply the prior $P(z,\vec{G}, \alpha)$ on all
the parameters simultaneously, but we do not have sufficient
observational constraints on the distributions of the physical
parameters and their redshift dependence.  Observationally
available constraints are often correlations between some limited combinations
of the parameters.  One instead could use semi-analytic models to
construct $P(z,\vec{G}, \alpha)$, but semi-analytic models do not
perfectly reproduce galaxy properties in the real universe and
the priors may well be biased.

What this  prior effectively does is to constrain the physical
parameters of the SPS models within realistic ranges.  
This motivates us to interpret the prior broadly and consider
constraining the parameter space by combining observed relationships
between physical properties of galaxies available in the literature
(which are basically in form of marginalized probability distribution functions).
We assume that the prior can be factorized by

\begin{eqnarray}
  P(z, \vec{G}, \alpha)=& P(z) P(SFR|M_*,z) P(\tau_V|SFR,z)\nonumber\\
  & P(age|M_*,z).
\end{eqnarray}

\noindent
This set of priors is not fully equivalent to $P(z,\vec{G}, \alpha)$,
but we are limited by the available observational information on
these physical parameters.
As mentioned above, metallicity is fixed to solar.  But, thanks to
the strong dust-age-metallicity degeneracy, the multi-color space is
covered by the dust and age variations.  Metallicity is difficult
to infer from broad-band photometry (e.g., \citealt{pforr12}) and
it is unlikely a useful prior in any case.
Because of the experimental aspect of the priors, we regard the analysis
in this paper as a proof-of-concept
analysis rather than fully contained work of physical priors.
We plan to further test and explore physical priors in the future.

The likelihood $P(\vec{m}|z,\vec{G}, \alpha)$ is computed as

\begin{equation}
  P(\vec{m}|z,\vec{G}, \alpha) \propto \exp \left( -\frac{\chi^2(z,\vec{G}, \alpha)}{2} \right),
\end{equation}

\begin{equation}
  \chi^2(z,\vec{G}, \alpha) = \sum_i \frac{(f_{i,obs}-\alpha f_{i,model}(z,\vec{G}))^2}{\sigma (f_{i,obs})^2},
\end{equation}

\noindent
where $f_{i,obs}$ and $f_{i,model}$ are observed and model fluxes in 
the $i$-th filter and $\sigma(f_{i,obs})$ is the observed flux uncertainty
in that filter.
We note that we use linear fluxes in the above equations, not logarithmic
magnitudes.  
We will later introduce a template error function, which assigns uncertainties
to model fluxes.  The above equations can be easily extended to add the model
uncertainties to the observed flux uncertainties in the quadrature.

Using Eqs. 4-6, we can now work out Eq. 3.  However, the marginalization
over $\alpha$ turns out to be a computationally expensive task.
We find that, in our cases discussed in the following sections,
$P(\vec{m}|z,\vec{G}, \alpha)$ is sharply peaked around $\alpha$ that
minimizes $\chi^2$ defined in Eq. 6, which can be computed by

\begin{equation}
  \frac{\partial \chi^2}{\partial \alpha}=0.
\end{equation}

\noindent
This can be explicitly written as

\begin{equation}
  \alpha_{best}=\frac{\sum_i \frac{f_{i,obs}f_{i,model}}{\sigma(f_{i,obs})^2}}
        {\sum_i \frac{f_{i,model}^2}{\sigma(f_{i,obs})^2}}.
\end{equation}

\noindent
Motivated by the sharp probability distribution function (PDF) around $\alpha_{best}$, 
we have compared photo-$z$'s computed with the marginalization over
$\alpha$ and those computed with $\alpha=\alpha_{best}$ without marginalization.
It turns out that they are almost identical: $\sigma(\Delta z_{phot})\sim0.001$
with no systematic offset between the two runs.  This level of scatter is much
smaller than those we discuss later and it does not affect our results at all.
Very faint objects
(e.g., those with photometric uncertainties larger than 0.5 mag in all filters)
may have a significant $P(\vec{m}|z,\vec{G}, \alpha)$ over a wide range of $\alpha$,
but photo-$z$ uncertainties for such faint objects are large in any case and
the extra uncertainty coming from assuming $\alpha=\alpha_{best}$ will be negligible.
For these reasons, we assume

\begin{eqnarray}
  P(z,\vec{G}|\vec{m}) & \propto & \int d\alpha P(\vec{m}|z,\vec{G},\alpha) P(z, \vec{G}, \alpha)\nonumber\\
  & \sim & P(\vec{m}|z,\vec{G},\alpha_{best}) P(z, \vec{G}, \alpha_{best}).
\end{eqnarray}

\noindent
A photo-$z$ PDF can then
be obtained by marginalizing over the model parameters:

\begin{equation}
  P(z|\vec{m}) \propto \sum_{\vec{G}} P(z,\vec{G}|\vec{m}).
\end{equation}

\noindent
Our choice of the model grids are described in the previous section.
A PDF is then normalized such that the probability integrated over
redshift equals unity.
In a similar way, we can compute PDFs for physical properties
(using $\alpha_{best}$ for SFR and stellar mass) with all the other
parameters marginalized over.  We shall emphasize that redshift is
also marginalized over and the uncertainty in redshift is
properly included in PDFs of physical properties.
These PDFs will be very useful for galaxy studies.

In the following, we first define the $N(z)$ prior and then move
on to describe the physical priors used to
constrain the parameter space of the SPS models.
One may rightly worry that the physical priors bias the inferred
physical properties.
We will quantify biases introduced by the priors in Section 8
and show that they are small.

%---------------------
\subsection{$N(z)$}

A redshift distribution of galaxies, $N(z)$, is often used as a prior
for photo-$z$'s in the literature and is probably the most popular one
(it is denoted as $P(z)$ in Eq. 4).
This prior is essentially an apparent, observed-frame luminosity function of
galaxies times unit volume integrated over redshift.  We could use a more
physically oriented prior such as rest-frame luminosity function or stellar
mass function.
After some experiments, it turned out that the $N(z)$ prior give similar
improvements to luminosity and stellar mass function priors.
Because $N(z)$ is expected to introduce less bias in the inferred physical
properties of galaxies, we choose to use $N(z)$.

Constraining $N(z)$ is not a trivial task because it requires unbiased
redshifts of a large number of objects down to faint enough magnitudes that we are
interested in.  One of the ways would be to use semi-analytic models as
done by \citet{brammer08}, although semi-analytic models may not perfectly
reproduce the $N(z)$ in the real universe \citep{henriques12}.
We here take an empirical approach and use very accurate photo-$z$ estimates
from \citet{ilbert09} to construct $N(z)$.  We note that \citet{ilbert09} did
not apply any $N(z)$ prior.

There are a number of functional forms of $N(z)$ adopted in the literature
(e.g., \citealt{benitez00,ilbert09,schrabback10,hildebrandt12}).  As discussed by
\citet{lefevre13}, $N(z)$ shows a high redshift tail at faint magnitudes.
We slightly tweak the functional forms adopted by these authors and take
the form of

\begin{equation}\label{eq:nz}
  N(z)\propto(z+0.1)^\alpha \left[ \exp \left[ -\left(\frac{z}{z_0}\right)^\alpha \right] +\beta\exp\left[ -\left(\frac{z}{z_0}\right)^{0.7\alpha}\right]\right]
\end{equation}

\noindent
where $z_0$, $\alpha$, and $\beta$ are free parameters.  This form
has fewer free parameters than those adopted in \citet{schrabback10}
and \citet{lefevre13}, but the high redshift tail is described well
by the 2nd exponent.  We also introduce a 'softening' parameter suggested by
\citet{hildebrandt12} to avoid too small probabilities at low redshifts,
which can be problematic for bright objects.

\begin{figure}[h]
\epsscale{1.}
\plotone{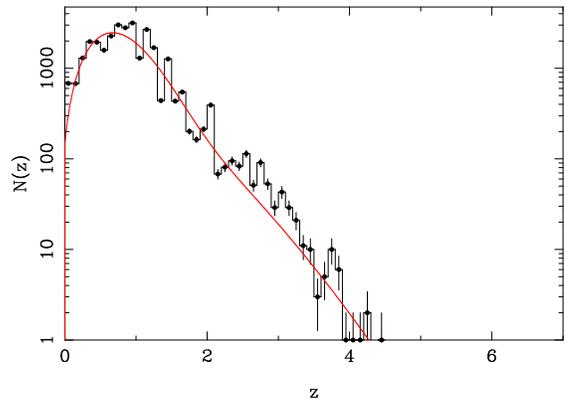}
\caption{
  $N(z)$ for galaxies with $23.0<i<23.5$. The histogram with points
  is the observed $N(z)$ and the curve is a fit to it.
}
\label{fig:dndz_prior}
\end{figure}

This prior can be extended
to include magnitude dependence, $P(z|m)$, as done by \citet{benitez00}.
We fit the functional form in Eq. \ref{eq:nz} to the photo-$z$ catalog
from \citet{ilbert09} in a number of $i$-band magnitude bins.
A sample fit is shown in Fig. \ref{fig:dndz_prior} and the fitted parameters
are summarized in Table \ref{tab:nz_param}.  We note that one of
the most popular codes {\sc bpz} applies a template-dependent $N(z)$ prior\footnote{
This template-dependent $P(z)$ prior constrains relative fractions of
galaxy populations (e.g., quiescent vs. star forming) as a function of redshift.
Therefore, a care is needed when it is used for galaxy population studies.
}, but we assume that this prior is independent of templates.

%---------------------
%---------------------

%---------------------------------
\begin{table}
  \begin{center}
    \begin{tabular}{cccc}
      $i$-mag range & $\alpha$ & $\beta$ & $z_0$\\\hline
      17.0--18.0 & 2.76 & 0.00 & 0.20\\
      18.0--19.0 & 2.89 & 0.10 & 0.26\\
      19.0--21.0 & 3.06 & 0.30 & 0.34\\
      21.0--21.5 & 2.32 & 0.00 & 0.53\\
      21.5--20.0 & 2.30 & 0.00 & 0.59\\
      20.0--22.0 & 2.26 & 0.00 & 0.44\\
      22.0--22.5 & 2.16 & 0.01 & 0.64\\
      22.5--23.0 & 2.07 & 0.06 & 0.69\\
      23.0--23.5 & 1.72 & 0.30 & 0.74\\
      23.0--24.0 & 1.82 & 0.15 & 0.71\\
      24.0--24.5 & 1.32 & 0.01 & 0.82\\
      24.5--25.0 & 1.33 & 0.01 & 0.88\\
      \hline
    \end{tabular}
  \end{center}
  \caption{
    Fitted parameters for $N(z|m_i)$.
  }
  \label{tab:nz_param}
\end{table}
%---------------------------------

%---------------------
\subsection{$P({\rm SFR}|M_*.z)$}

There is a well-known correlation between SFR and stellar mass (e.g.,
\citealt{daddi07,elbaz07,wuyts11b,whitaker12}).  This relation has
been extensively studied across redshifts and we use it as a priori
knowledge to constrain the model templates.  The relation between SFR
and stellar mass is not quite linear, but for simplicity, we assume a
linear correlation between SFR and stellar mass $M_*$;

\begin{equation}\label{photoz:eq:sfr_star}
  {\rm SFR}_{\rm SF}(M_*, z)={\rm SFR}^*(z)\times \frac{M_*}{10^{11}{\rm M_\odot}}\rm\ M_\odot\ yr^{-1},
\end{equation}

\noindent
where

\begin{equation}\label{photoz:eq:sfr_star2}
  {\rm SFR}^*(z)= \left \{
  \begin{array}{ll}
    10\times(1+z)^{2.1}  & (z<2),\\
    19\times(1+z)^{1.5} & (z\geq2).
  \end{array}\right.
\end{equation}

\noindent
At $z=0$, a typical SFR of a star forming galaxy with $10^{11}\rm M_\odot$ is
$10\rm M_\odot\ yr^{-1}$ and SFR at fixed mass increases at higher redshifts
such that SFR is 10 times larger at $z=2$.  At higher redshifts,
there is currently no consensus about the evolution of the
star formation sequence.  \citet{gonzalez10} suggested that
the evolution of sSFR (which is proportional to SFR at fixed mass)
flattens out at $z>2$, while \citet{stark13} hinted at a gradual
increase at higher redshifts.  We assume a gradual increase
shown in \citet{stark13} with a redshift dependence of $(1+z)^{1.5}$.
We have confirmed that there is no strong effect on photo-$z$
accuracies if we assume no evolution at $z>2$ for the photo-$z$
tests performed in Section 6.

While star forming galaxies show a strong correlation between
SFR and stellar mass, quiescent galaxies are off from that relation.
We define a sequence for quiescent galaxies as the star forming 
sequence offset by $-2$~dex.  SED fitting does not reproduce  SFRs at
very low SFRs \citep{pacifici12} and we find that this form works
well after some experiments.  SFRs can numerically be very small
(e.g., $10^{-9}\rm\ M_\odot\ yr^{-1}$) and SFRs lower than $-2$ dex from
the star forming sequence are forced to be $-2$ dex just for
computational reasons.

\begin{figure}[h]
\epsscale{0.8}
\plotone{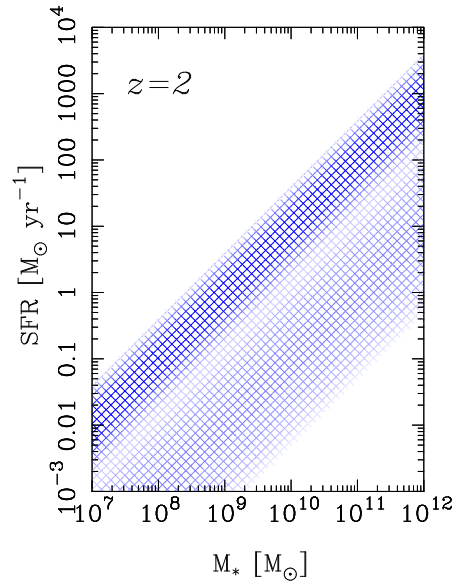}
\caption{
  Schematic plot of SFR vs stellar mass prior at $z=2$.  A darker color
  means a higher probability.  Note that this prior is redshift dependent
  and the prior at $z=0$ for instance is offset by $-1$~dex in SFR.
}
\label{fig:sfr_smass_prior}
\end{figure}

Putting all this together, we assume that galaxies exhibit
the two distinct sequences formed by star forming and quiescent
galaxies and a probability of finding a galaxy with a star formation
rate of SFR with a given stellar mass at redshift $z$ can be
expressed as a sum of two Gaussians:

\begin{eqnarray}
&&  P({\rm SFR}|M_{*},z)\propto\nonumber\\
&&  \frac{1}{\sigma_{\rm SF}}\exp\left[-\frac{1}{2}\left(\frac{\log {\rm SFR}-\log {\rm SFR}_{\rm SF}(M_*,z)}{\sigma_{\rm SF}}\right)^2\right]\nonumber\\
&&  +\frac{1}{\sigma_{\rm Q}}\exp\left[-\frac{1}{2}\left(\frac{\log
      {\rm SFR} - \log {\rm SFR}_{\rm SF}(M_*,z)+2}{\sigma_{\rm Q}}\right)^2\right]\nonumber,\\
&&
\end{eqnarray}

\noindent
where ${\rm SFR}_{\rm SF}(M_{*},z)$ is the mean SFR
for a star forming galaxy at redshift $z$ and with stellar mass
$M_{*}$ defined in Equation~(\ref{photoz:eq:sfr_star}).
We assume the relative fraction between the two populations
is 1:1 at all redshifts to be conservative.
$\sigma_{\rm SF}$ and $\sigma_{\rm Q}$ are the dispersions in each of
the star forming and quiescent sequence.  Here we take
$\sigma_{\rm SF}=0.3$~dex and $\sigma_{\rm Q}=1$~dex.  The observed
scatter of the sequence may be smaller (e.g., 0.2 dex; \citealt{daddi07}),
but starbursting galaxies are often located off from the sequence \citep{rodighiero11}
and we choose to assume a slightly large scatter.  As for the quiescent
sequence, low SFRs are difficult to measure from SEDs \citep{pacifici12}
and we adopt a large scatter.  But, we should emphasize that
photo-$z$ accuracies are not very sensitive to the choice of
the scatters within reasonable ranges.
Fig. \ref{fig:sfr_smass_prior} schematically illustrates this prior.

%---------------------
%---------------------

%---------------------
\subsection{$P(\tau_V|SFR,z)$}

An amount of attenuation is known to positively correlate with SFR
as well as stellar mass (e.g., \citealt{hopkins03,garn10}).
This fact has often been employed in photo-$z$ computations
in the literature. E.g., \citet{fontana04} limited the attenuation to
$A_V<0.6$ for low-SFR galaxies with $age/\tau\geq4$, and \citet{ilbert09}
did not apply extinction to their quiescent templates.  Here, we
formulate the relationship and implement it as a prior.

\citet{sobral12} suggested that the relationship between attenuation
and stellar mass seems to remain unchanged out to $z\sim1$.
One might use this interesting relation as a prior, but the problem is
that the relation is actually bimodal --- at a given stellar mass,
there are star forming and quiescent populations and only
the former suffers from attenuation.  To avoid this bimodality,
we  use the redshift-dependent SFR-attenuation relation\footnote{
The redshift independence of the mass-attenuation relation 
might appear odd at a first glance given the evolving
relationship between SFR and stellar mass discussed above.
The evolving SFR-mass relation seems to be compensated by the evolving
SFR-attenuation relation, making the mass-attenuation independent
of redshift.
}.
We define an evolution corrected SFR as

\begin{equation}
  {\rm SFR}_0=100\frac{{\rm SFR}}{{\rm SFR}^*(z)},
\end{equation}

\noindent
where ${\rm SFR}^*(z)$ is defined in Eq. \ref{photoz:eq:sfr_star}
and is used to eliminate the redshift dependence of the SFR-attenuation
relation here.  
A factor of 100 is motivated to set the threshold SFR above which 
$\tau_V$ correlates with SFR to unity.
We adopt a Gaussian form for this prior:

\begin{equation}
P(\tau_V | {\rm SFR},z)\propto
  \exp \left[ -\frac{1}{2}\left(
    \frac{\tau_V-<\tau_V>}{\sigma_{\tau_V}}\right)^2\right],
\end{equation}

\noindent
where

\begin{equation}
<\tau_V>=\left \{ \begin{array}{ll}0.2 & ({\rm SFR}_0<1),\\
  0.2+0.5\log {\rm SFR}_0 & ({\rm SFR}_0>1).\end{array}\right.
\end{equation}

\begin{figure}[h]
\epsscale{1.}
\plotone{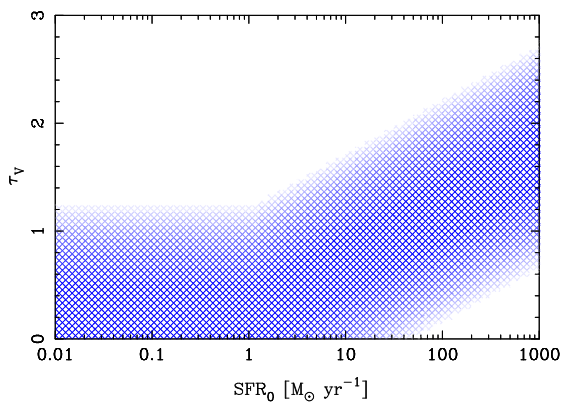}
\caption{
  Schematic plot of $\tau_V$ vs SFR$_0$ prior.  SFR$_0$ is a function
  of redshift.  For reference, SFR$_0=1\rm\ M_\odot\ yr^{-1}$ corresponds
  to SFR$=0.1\rm\ M_\odot\ yr^{-1} $ at $z=0$ and SFR$=1\rm\ M_\odot\ yr^{-1}
  $ at $z=2$. 
}
\label{fig:tauv_sfr0_prior}
\end{figure}

\noindent
If a galaxy has a low SFR with its normalized SFR less than unity,
the mean attenuation is fixed to a small value of 0.2.  This non-zero
value is adopted in order to leave room to compensate for metallicity
variations.  But, we have confirmed that photo-$z$ accuracies do not
significantly change if we set it to zero.
For the same reason, we adopt a relatively large $\sigma_{\tau_V}$ of 0.5.
The threshold $\rm SFR_0$ is equivalent to SFR=$0.1\rm M_\odot\ yr^{-1}$
at $z=0$ and is motivated by \citet{garn10}.  At higher SFRs, the mean
attenuation increases with SFR.  The adopted functional form
reproduces the observation by \citet{garn10}.
The prior is schematically illustrated in Fig. \ref{fig:tauv_sfr0_prior}.

We should note that the attenuation that \citet{garn10}
measured is for H$\alpha$, which is known to suffer from a larger
attenuation than the stellar continuum \citep{calzetti97}.
The $\tau_V$ parameter in our models defines attenuation for
stellar continuum 
and thus the mean attenuation assumed above should have been smaller
by about a factor of 2.  However, we find that the shallower relation results in
poorer photo-$z$ accuracies.  We speculate that this is due to a combined
effect of (a) assumption of solar metallicity for all models and (b)
dependence of attenuation curve on SFRs.  As mentioned above, 
there is degeneracy between metallicity, age, and attenuation.
We fix the metallicity to solar and let age and attenuation vary to
compensate for the metallicity variation in real galaxies, which
may affect the SFR-attenuation relation.  Also, as discussed by
\citet{ilbert09}, the attenuation curve may be dependent on SFR,
which also affects the relation.
This is obviously an area where further work is needed.

%---------------------
%---------------------

%---------------------
\subsection{$P(age|M_*,z)$}

It is known, from experience, that templates with very young ages often
give poor redshift accuracies and/or inaccurate physical properties
(e.g., \citealt{fontana04,pozzetti07,tanaka10,wuyts11a}).
This is primarily due to degeneracy introduced by these young
templates; young templates all look similar regardless of their star formation
timescales and they look similar even to older templates with long
star formation timescales when only a small number of filters are available.
All this is because the overall spectrum is dominated by young stars.
Very young templates and/or templates
with short $\tau$ are often manually excluded to reduce the degeneracy
in the literature.
Rather than manually removing them from the library,
we prefer to put it on a more physical ground.

A crude, time-averaged mean star formation rate of a template can be given by

\begin{equation}
  {\rm <SFR>}=M_*/age.
\end{equation}

\noindent
This is not strictly a mean SFR because the stellar mass-loss is not accounted for
but is a reasonable proxy.  Except for starbursting galaxies, steady-state
star forming galaxies (i.e., those on the star formation sequence) have
SFRs of up to a few hundred $\rm M_\odot\ yr^{-1}$.  Those are the most
massive galaxies whose stellar mass (or equivalently SFR here) function
follows an exponential form.    Motivated by this, we assume a function of

\begin{equation}
  P(age|M_*,z)\propto\exp\left[-\frac{\rm <SFR>}{{\rm SFR}^*(z)}\right].
\end{equation}

\begin{figure}[h]
\epsscale{1.0}
\plotone{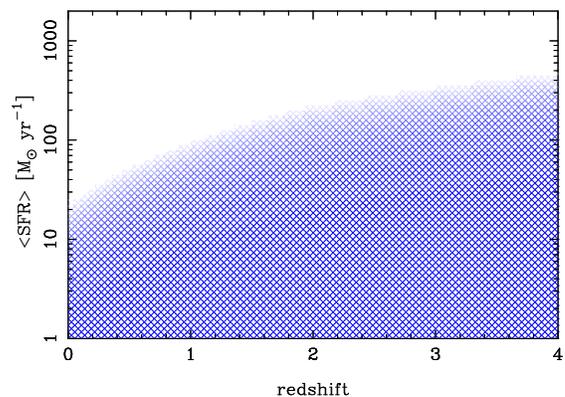}
\caption{
  Schematic plot of the $<SFR>$ vs redshift prior.
}
\label{fig:meansfr_z_prior}
\end{figure}

\noindent
This form gives a low probability to young templates and is effectively
similar to the hard-cut priors adopted in the previous studies.  A notable
difference is that this form is more effective for more massive galaxies;
a massive galaxy with young age has a low probability, while a low-mass galaxy
with the same age has a high probability.  This makes sense because
low-mass galaxies are expected to be younger than massive galaxies.
Another difference is that the characteristic SFR evolves with redshift in
an observationally motivated way.  This prior is schematically shown in
Fig. \ref{fig:meansfr_z_prior}.

%---------------------
%---------------------

%--------------------------------------------------------------------
\subsection{Probability distribution function}

The code produces a redshift PDF and
the full PDF information should be used for any science analysis.
But, it may be useful to make a point estimate for
each object for certain purposes.  A PDF of a galaxy can be quite complex
and one faces a question of how to make such a point estimate.  The most
popular ways include:  (1) peak of the PDF, (2) weighted mean, and (3) the median of the PDF.
The definition of the first one is obvious.  The latter two are defined as

\begin{equation}
<z>=\int^{z_{max}}_{z_{min}} z P(z) dz,
\end{equation}

\noindent
and

\begin{equation}
\int^{z_{median}}_{z_{min}} P(z) dz=0.5,
\end{equation}

\noindent
where $z_{min}$ and $z_{max}$ are 0 and 7 in our case.  After some experiments,
we find that the median redshift works best.  One might expect that the weighted
redshift is a reasonable method because it uses the full PDF.
But, a PDF often has multiple peaks and, depending on the relative
probabilities of the peaks, the weighted redshift often lies in
between the peaks.
One can imagine a simple case in which there are two peaks with the same probability
in a PDF, and the weighted redshift will be exactly in the middle of the peaks,
where there is little probability.

The peak redshift is often adopted in the literature, but 
the median redshift works better.  It is likely because the median redshift
is a better tracer of the 'primary' peak in PDF.
Peaks in a PDF have different heights and widths
and the peak with the largest {\it integrated} probability, not the {\it peak}
probability, is more likely to be a correct redshift.  This primary peak
could be better identified by integrating the probability as done in
the median redshift, not by locating the highest peak.  One can explore
an algorithm to directly identify the primary peak rather than taking the median
of PDF, but for computational simplicity, we use the median redshifts in this paper.

%-------------------------------------------------------------------
%-------------------------------------------------------------------
\section{Template error function}

We use model templates generated with an SPS code.
These model templates are subject to systematic uncertainties.
Such uncertainties include deviation of star formation histories
of real galaxies from the simple $\tau$-model, ignorance of
metallicity distribution, and errors in the stellar
evolutionary track (which introduces an age-dependent systematics), etc. 
These systematic uncertainties cause mismatches between model SEDs
and observed SEDs of galaxies.  We use a template error function
to crudely correct for such systematics.
The template error function is first introduced by \citet{brammer08} and 
it is used to assign random uncertainties to templates.  But, we
extend it to include correction for systematic flux errors.

We define the template error function with two terms;
systematic flux stretch and random flux uncertainty.  The first term is a flux
correction to model templates as a function of rest-frame wavelength
and is meant to reduce mismatches with observed SEDs.
This correction is typically small, of order a few percent, but
can be as large as 10 per cent at some wavelengths.
The second term assigns random uncertainties to the model fluxes
to weight reliable parts of the SEDs, which is
reasonable because model templates are often calibrated in
the rest-frame optical and they can be more uncertain at other
wavelengths.

A template error function can be easily coupled with zero-point
offsets in photometric data in real analysis.
In order to separate these two components, we first correct for
systematic zero-point offsets by fitting spectroscopic objects
located at low redshifts with their redshifts fixed to their
spectroscopic redshifts.  Given that the Sloan Digital Sky
Survey \citep{york00} provides a large number of galaxies at
$z\lesssim0.2$ over a wide area of the sky, we use objects located
at $z<0.2$ and take the median of $f_{obs}/f_{model}$ to estimate
the zero-point offsets in a given filter.  This process normally converges
(i.e., offsets become less than 1\%) after a few iterations.
Note that no template error functions are applied in this process.

We then construct a template error function.  For this, we need a large
sample of spectroscopic redshifts over a wide range of redshift.
This is fairly demanding and there are only a few deep fields that
allow us to make error functions using spec-$z$'s (e.g., Chandra
Deep Field South as done in \citealt{tanaka13a}).
A practical alternative would be to use very accurate
photo-$z$'s computed with many bands, which can be as accurate as
$\sigma (\Delta z/(1+z))\sim0.01$, for this job.
We use the $ugriz$ photometry and the precise photometric redshift from
\citet{ilbert09} to construct
template error functions.  We do not include medium-bands nor
narrow-bands here because we are interested in flux
errors smoothed over a typical bandwidth
of broad-band filters.  Narrow/Medium-band filters will give 
flux stretches on a finer wavelength scale.  In other words,
if one would like to apply a template error function to
medium-band filters, that function has to be estimated separately
from broad-bands.

We compute the systematic differences between the observed
fluxes and the best-fit model fluxes using objects
with precise photometric redshifts.  Because objects spread over
a range of redshift, we can effectively cover a wide range of
rest-frame wavelength with the broad-band photometry alone.
The median of $f_{obs}/f_{model}$ at a given rest-frame
wavelength is used to correct for the systematic template errors
(1st term) and
dispersion around the median is used as an uncertainty of
the model templates at that rest-frame wavelength (2nd term).
A typical
photometric uncertainty in the observed data is subtracted off
from the measured dispersion in order to derive intrinsic
uncertainties in the models.
In principle, one could apply such a correction to templates of a given
spectral type. 
But, for now we simply
make several redshift bins and apply a single ``master''
correction to all the templates in each bin in this paper.

\begin{figure}[h]
\epsscale{1}
\plotone{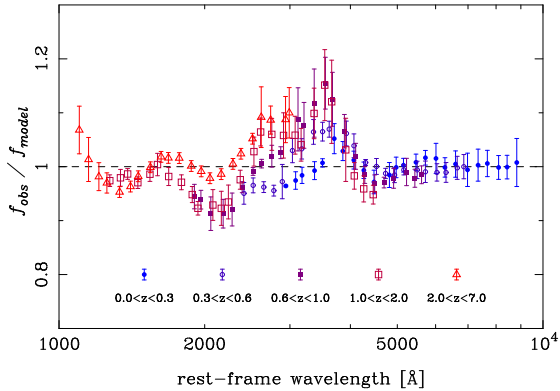}
\caption{
  Template error functions in several redshift bins.
  The points show flux stretches and
  the error bars show intrinsic uncertainties in the model fluxes.
  The symbols are explained in the figure.
}
\label{fig:errfn}
\end{figure}

Fig. \ref{fig:errfn} shows the template error functions in several
redshift bins generated
with the COSMOS data.  The corrections are typically a few percent,
but there are a few specific rest-frame wavelength ranges where
the corrections can be as large as 10\%.  \citet{maraston09} showed
that the observed $g-r$ color of passively evolving galaxies 
at $z\sim0.4$ is bluer by about 0.1 mag compared to models with
an empirical stellar spectral library.  Although a direct comparison
with our template error functions cannot be made because we include
star forming galaxies as well here, it is consistent
that models are too red around the rest-frame 4000\AA\ by about 10\%.
It is interesting to note that the template error functions evolve
with redshift; the correction at $\sim3500\rm\AA$ increases towards
$z\sim1$.  This has an implication for an age-dependent systematics 
in the SPS models, but a detailed study of it is beyond the scope
of this paper.  A template error function may provide an interesting
way to address issues with SPS models using photometric data.

%---------------------
%---------------------

%-------------------------------------------------------------------
%-------------------------------------------------------------------
\section{Improvements in photometric redshifts}

We move on to demonstrate how the physical priors and template error functions
improve photo-$z$'s.  We first describe the input photometric data and
then define quantities to characterize photo-$z$'s.  After that, we
present our photo-$z$'s and discuss some of the outstanding issues.

A photo-$z$ accuracy is dependent on available filters and 
we focus on $griz$ and $ugriz$ photometry in the rest of the paper
because these are the filter sets often available in large surveys;
the HSC survey and DES observe in $grizy$, CFHTLenS (which will be
discussed in the next section) in $ugriz$, and LSST in $ugrizy$.
The $y$-band photometry is not available in the public COSMOS data
and thus the photo-$z$ accuracy we discuss below is a lower
limit of these surveys.
Another reason for this limited filter set is because priors play
an important role.
Many filters over a wide wavelength range provide a strong constraining
power on SED shapes and redshifts and priors become less important
(e.g., \citealt{ilbert09} did not apply any priors because they had
the 30-band photometry).  Priors are more effective when less photometric
information is available and the $(u)griz$ photometry is a good
starting point to investigate how effective the physical priors introduced
in Section 4 are.

%----------------
\subsection{Input photometric catalog}

Obvious sources to characterize the accuracy of photo-$z$'s are
spectroscopic redshifts.  There are a number of flux-limited/color-selected
spectroscopic surveys in the literature.  Some of the largest ones would
include SDSS \citep{york00}, GAMA \citep{driver11}, VVDS \citep{lefevre05},
VIPERS \citep{guzzo13}, PRIMUS \citep{coil11}, DEEP2 \citep{davis03}, and zCOSMOS \citep{lilly07}.
The deepest of these is VVDS-UltraDeep, which is a flux limited survey at
$23.0<i<24.75$.  It is a very unique source of absorption line redshifts of 
high redshift galaxies, which are mostly missed from shallower surveys.
Unfortunately, however, even this deepest survey is still not deep enough
to calibrate and validate photo-$z$'s in the on-going/upcoming surveys, which
reach $i\sim26$ or deeper.  There is currently no spectroscopic sample that
allows us to calibrate photo-$z$'s down to this faint magnitude.
Imaging goes deeper than spectroscopy.

A practical calibration ladder to reach faint magnitudes would be to
use many-band photo-$z$'s often available in deep fields.
One notable example of this is photo-$z$'s available in COSMOS \citep{ilbert09},
and in this section, we compare our $griz$ photo-$z$'s with
the 30-band photo-$z$'s down to $i=25.0$.
This is not as deep as we would like, but the public catalog does not
contain fainter galaxies.  Also, the COSMOS data are probably not deep
enough to obtain good photo-$z$'s at fainter magnitudes.
The catalog is $i$-selected and is suited to test our code in
the context of large surveys, in which objects are often detected 
in the $i$-band for weak-lensing science.
There are more recent near-IR selected catalogs \citep{ilbert13,muzzin13},
but most of the above mentioned large surveys are optical surveys
and we stick with the $i$-selected catalog.
We note that the effects of the physical priors and template error functions
demonstrated below   are not dependent on the detection filter.
We show in the Appendix that they indeed significantly improve
photo-$z$'s for a $K$-selected catalog.

We add 0.02 mag uncertainty in the quadrature to the cataloged uncertainties
in \citet{ilbert09} because the uncertainties seem to be underestimated
(reduced $\chi^2$ are often large for bright objects).
This affects only bright galaxies for which uncertainties are smaller than $\sim0.02$~mag.
We also note that \citet{ilbert09} increased the uncertainties by a factor of 1.5
(which affects all galaxies) for their photo-$z$ analysis.

%---------------------------------
\begin{table*}[htb]
  \begin{center}
    \begin{tabular}{cccc|ccccc}
      case & physical priors & $N(z)$ prior & template errfn   & bias     & $\sigma$& $f_{outlier}$ & $\sigma_{conv}$ & $f_{outlier,conv}$\\\hline
      1    & No              & No           & No               & $+0.011$ & $0.143$ & $30.9\%$    & $0.182$       & $45.2\%$\\
      2    & Yes             & No           & No               & $-0.017$ & $0.101$ & $24.2\%$    & $0.112$       & $30.2\%$\\
      3    & Yes             & Yes          & No               & $-0.027$ & $0.094$ & $19.7\%$    & $0.100$       & $24.7\%$\\
      4    & Yes             & Yes          & Yes (w/o offset) & $-0.019$ & $0.082$ & $20.6\%$    & $0.087$       & $22.2\%$\\
      5    & Yes             & Yes          & Yes (w/  offset) & $-0.001$ & $0.084$ & $20.8\%$    & $0.081$       & $22.9\%$\\
      6    & No              & Yes          & Yes (w/  offset) & $+0.004$ & $0.101$ & $19.1\%$    & $0.095$       & $25.8\%$\\
      \hline
    \end{tabular}
  \end{center}
  \caption{
    Summary of photo-$z$ accuracies.
  }
  \label{tab:photoz}
\end{table*}
%---------------------------------

%----------------
\subsection{Quantities to characterize photo-$z$}

There are a few standard quantities used to characterize photo-$z$ accuracies;
bias, dispersion, and outlier rate.  The definitions adopted in the literature
are not always the same and we explicitly define them here for this work.

\begin{itemize}

\item {\bf Bias:}
Photo-$z$'s may systematically be off from spectroscopic redshifts
and we call this systematic offset bias.  We compute a systematic bias
in $(z_{\rm phot}-z_{\rm spec})/(1+z_{\rm spec})$ by applying the biweight statistics
\citep{beers90}.
We iteratively apply 3$\sigma$ clipping for 3 times to reduce outliers.

\item {\bf Dispersion:}
In the literature, dispersion is often computed as

\begin{equation}
\sigma_{\rm conv}=1.48\times{\rm MAD}\left(\frac{z_{\rm phot}-z_{\rm
    spec}}{1+z_{\rm spec}}\right),
\end{equation}

\noindent
where MAD is the median absolute deviation.  Note that this definition
does not account for the systematic bias.  In addition to this conventional
definition, we also measure the dispersion by accounting for the bias
using the biweight statistics.  We iteratively apply a $3\sigma$ clipping
as done for bias to measure the dispersion around the central value.
 We denote the conventional dispersion
and the biweight dispersion as $\sigma_{\rm conv}$ and $\sigma$, respectively.

\item {\bf Outlier rate:}
The conventional definition is

\begin{equation}
f_{\rm outlier,conv}=\frac{N\left(\frac{|z_{\rm
      phot}-z_{\rm spec}|}{1+z_{\rm spec}}>0.15\right)}{N_{\rm total}},
\end{equation}

\noindent
where outliers are defined as $|z_{\rm phot}-z_{\rm spec}|/(1+z_{\rm spec})>0.15$.
Again, this definition does not account for the systematic bias.
The threshold of 0.15 is an arbitrary value but is probably fine for
photo-$z$'s with several bands.  It is clearly too large for those with many bands.
Together with this conventional one, we also define outliers as those
$2\sigma$ away from the central value (these $\sigma$ and center are
from biweight; see above). This $2\sigma$ is an arbitrary choice, but
it is motivated to match reasonably well with the conventional
one for several band photo-$z$'s.  We will denote the $\sigma$-based
outlier fraction as $f_{\rm outlier}$ and the conventional one as
$f_{\rm outlier,conv}$.
\end{itemize}

%----------------
\subsection{Photo-$z$ vs. spec-$z$}

%---------------------
\begin{figure*}
\plottwo{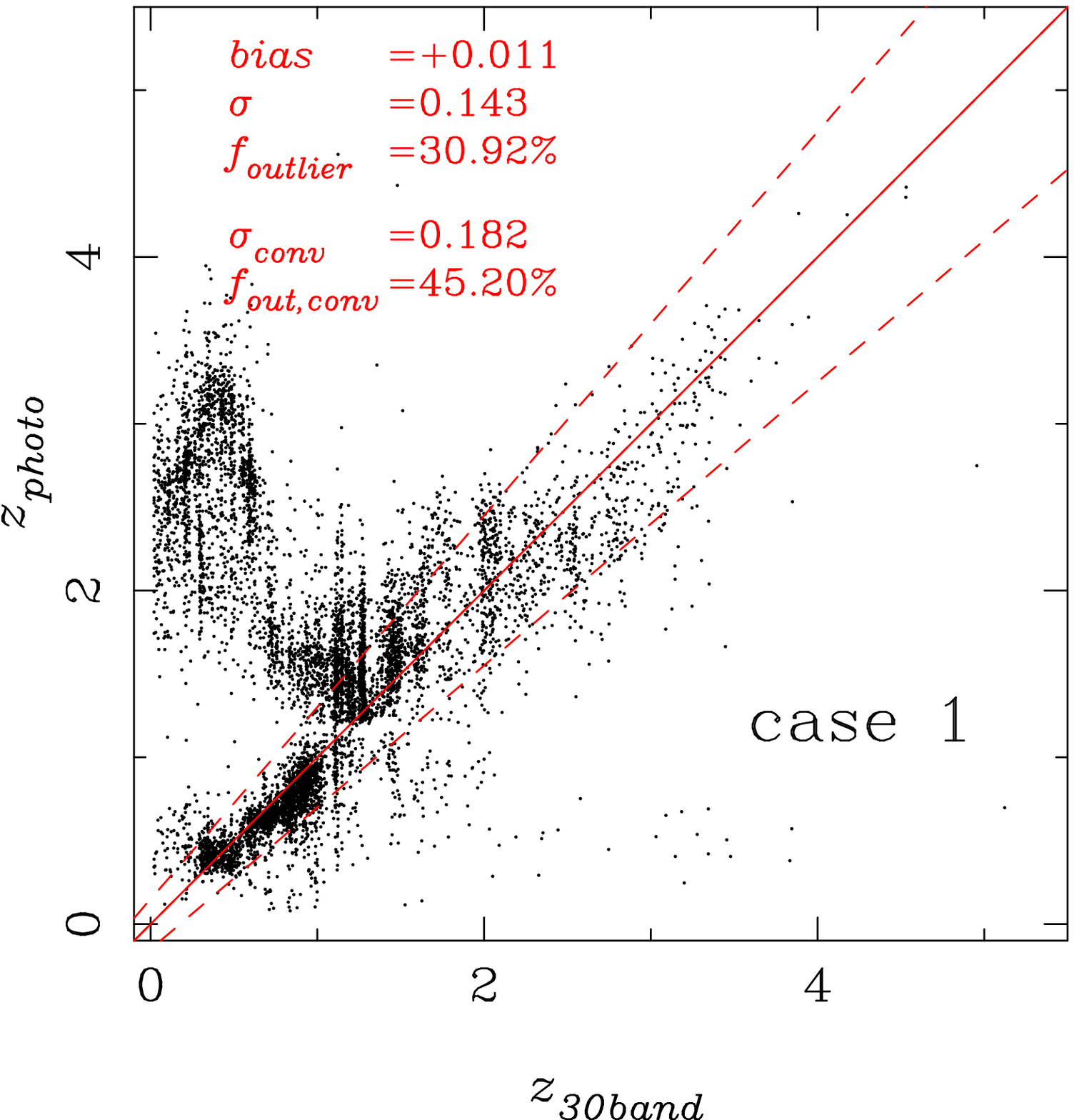}{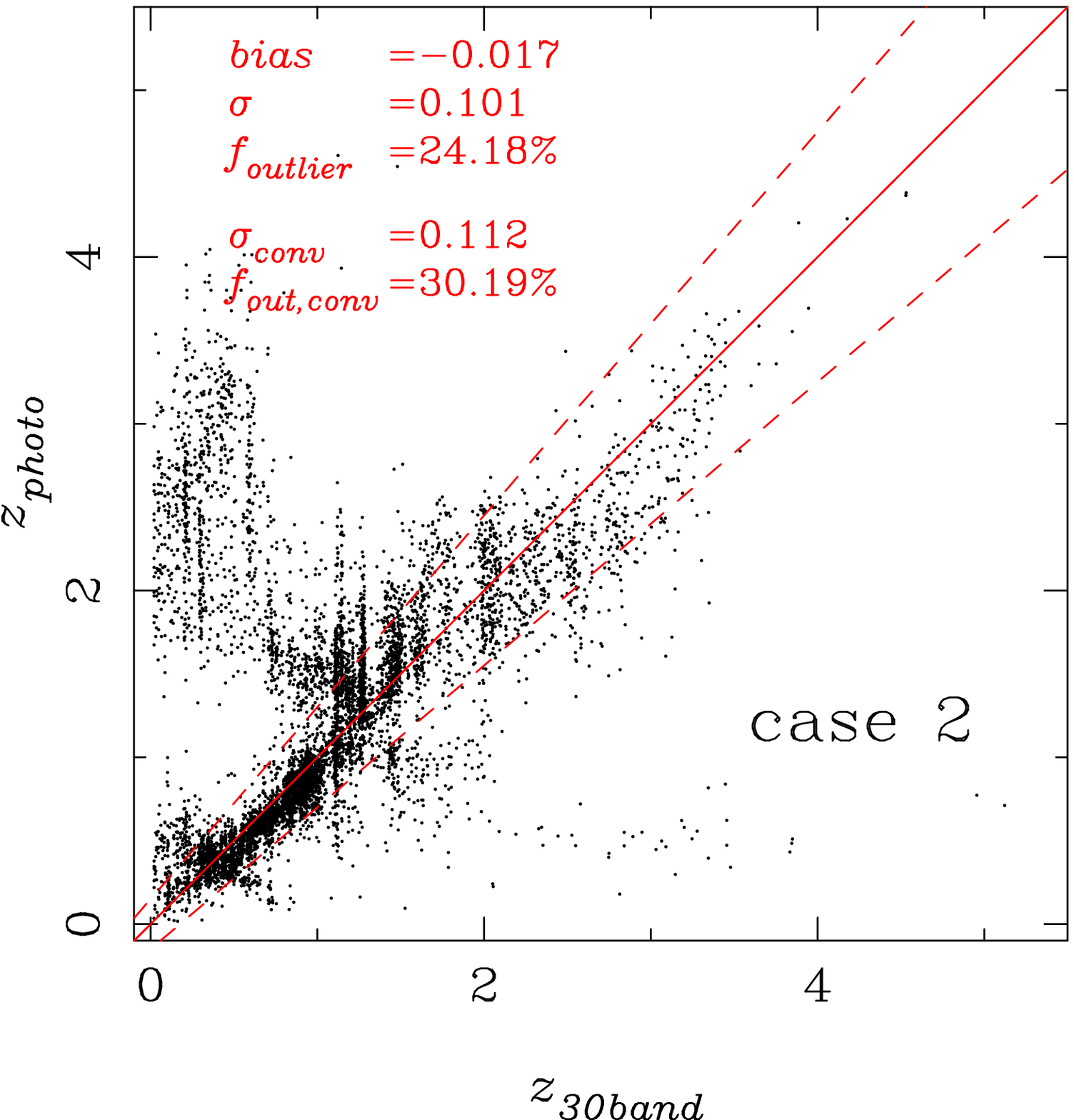}%\vspace{0.5cm}
\plottwo{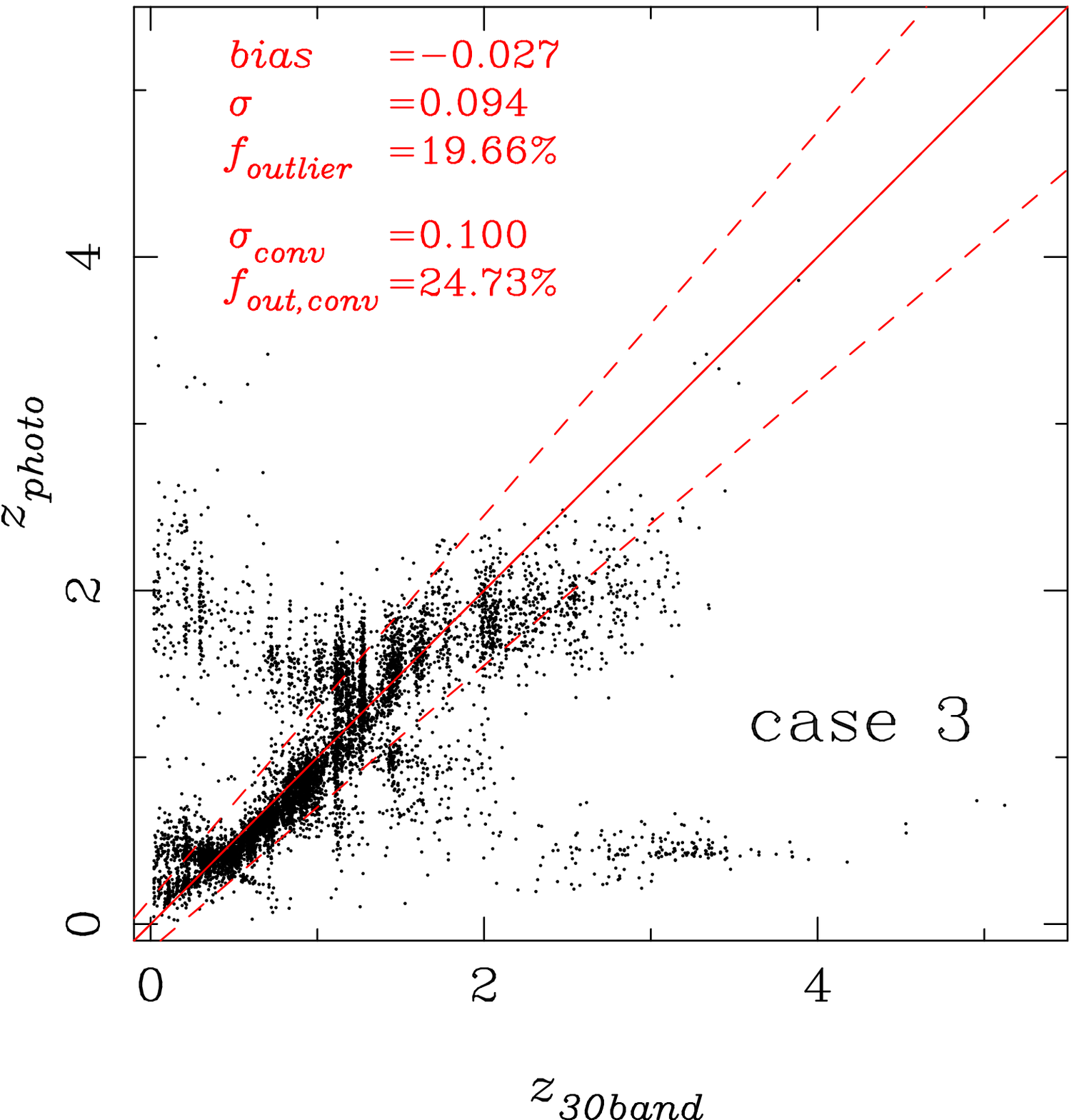}{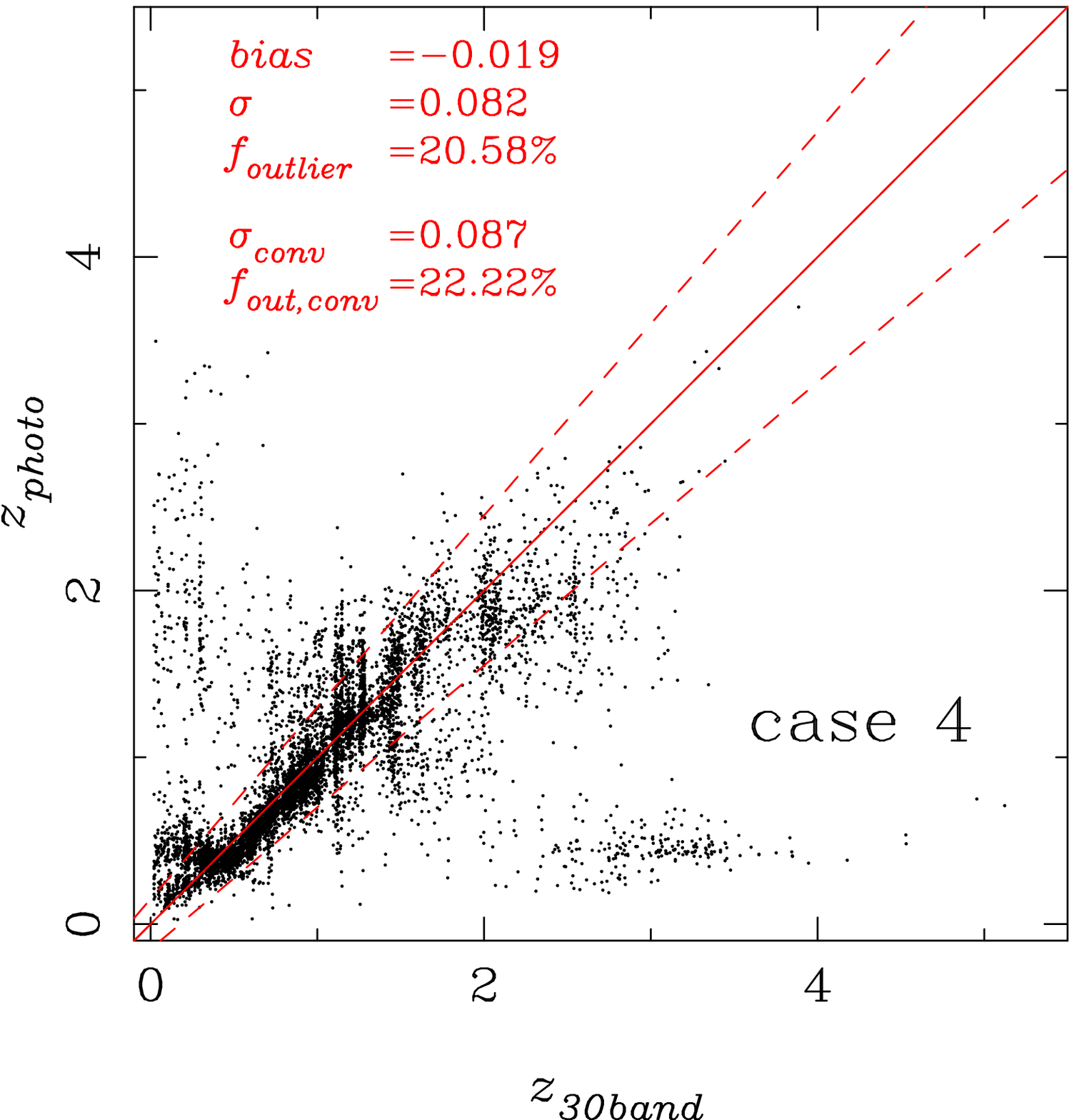}%\vspace{0.5cm}
\plottwo{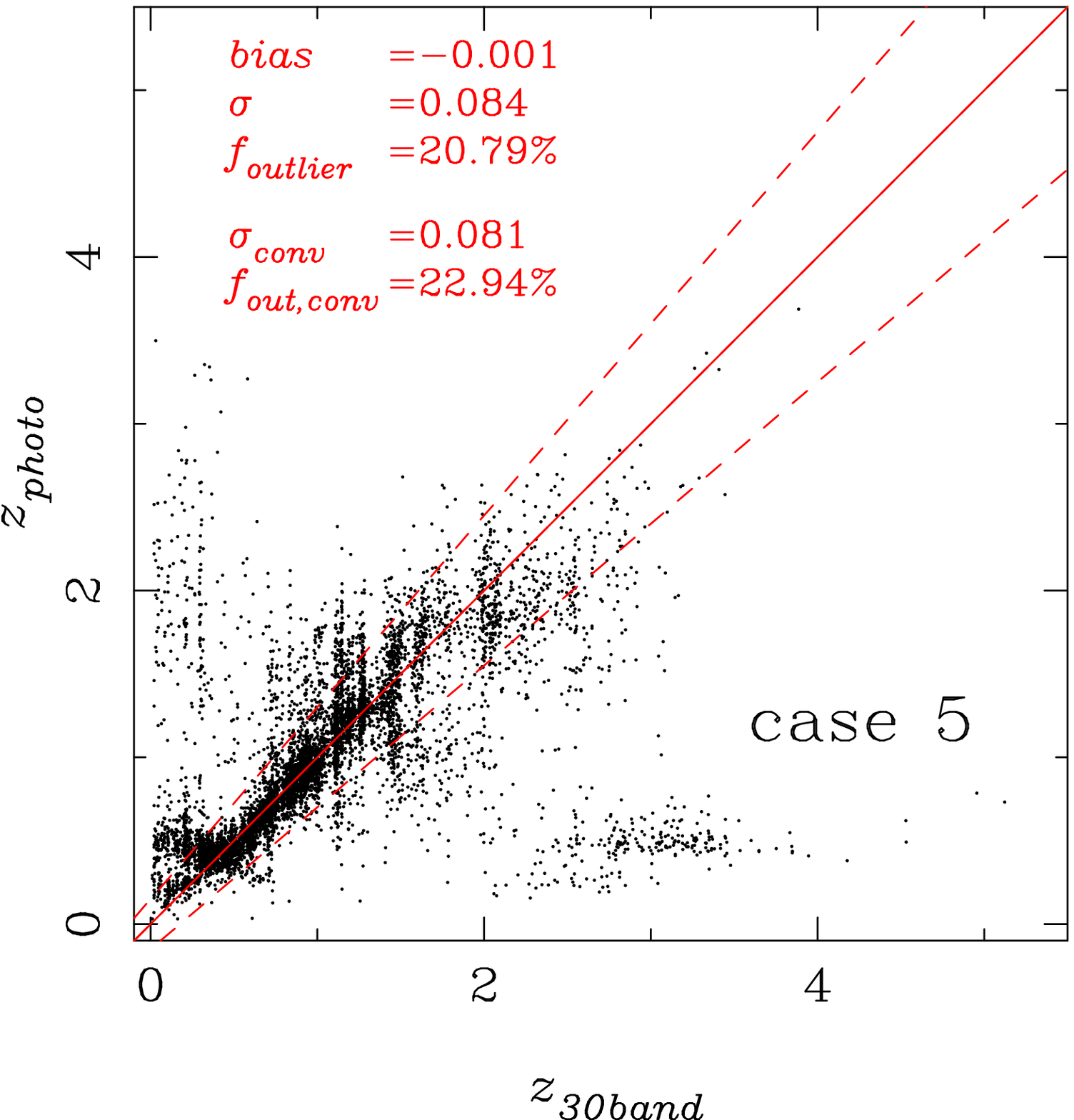}{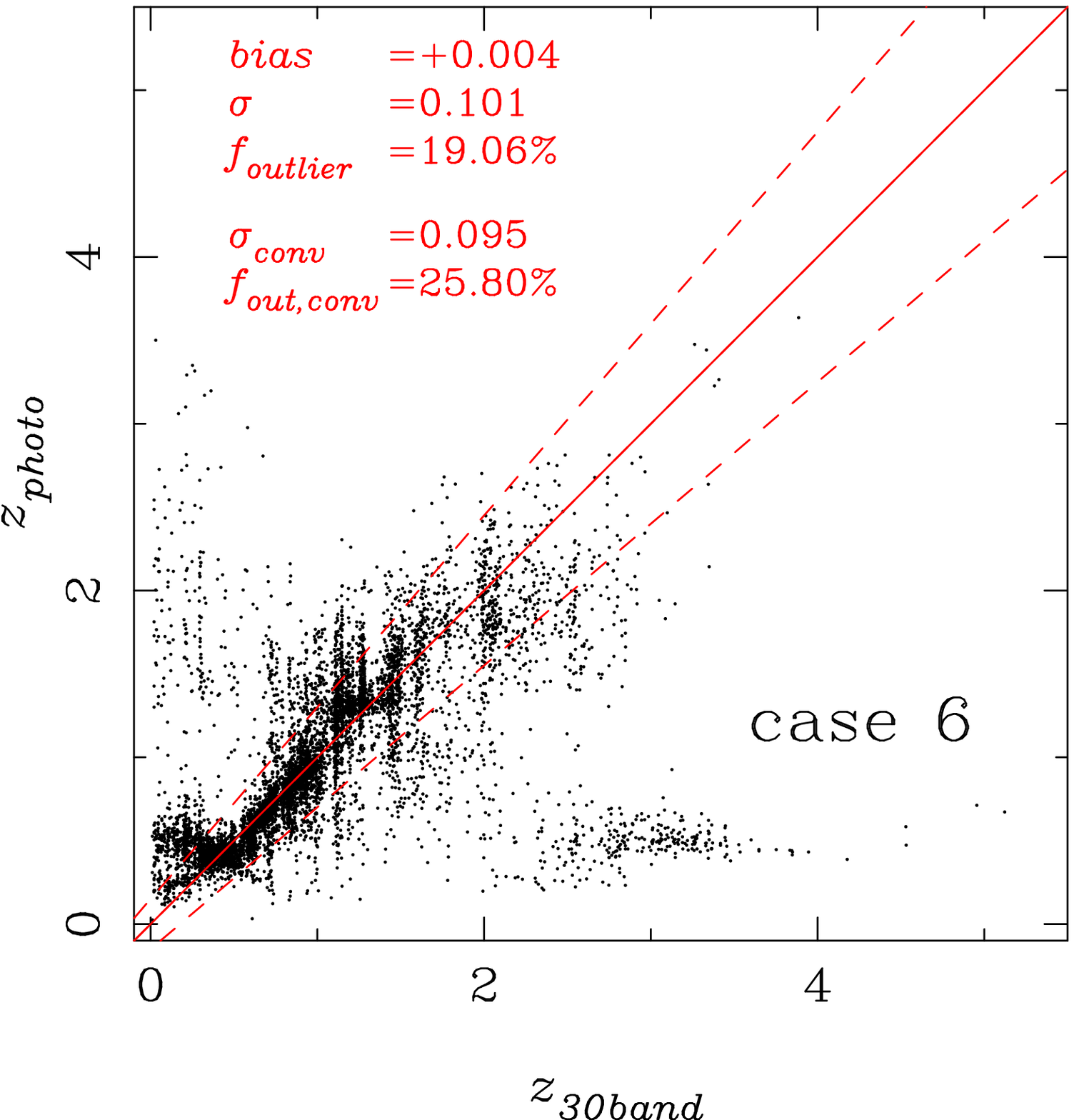}
\caption{
  $z_{phot}$ plotted against $z_{30band}$.  The solid line is $z_{phot}=z_{30band}$
  and the dashed lines show $|z_{phot}-z_{30band}|/(1+z_{30band})=\pm 0.15$.
  Objects above/below these lines are regarded as outliers in the conventional
  definition.  The plots are for 6 cases defined in Table \ref{tab:photoz}.
  Objects with $i<25$ are included and every 30 objects are plotted for clarity.
  No object clipping is applied in the statistics here.
}
\label{fig:cosmos_photoz}
\end{figure*}
%---------------------

Let us now demonstrate how the priors and template error function improve
photo-$z$.  Table \ref{tab:photoz} summarizes the combination of the priors
and template error functions and the resultant photo-$z$ accuracies
discussed in this section.  We shall emphasize
that some of the many-band photo-$z$'s that we compare with may be
incorrect and thus we should not over-interpret the numbers discussed below
in the absolute sense.

We start with the top-left panel of Fig. \ref{fig:cosmos_photoz} (case 1),
in which
photo-$z$'s computed without any of the physical priors and template
error functions are compared with the high-accuracy photo-$z$'s.
As indicated in the plot, the scatter and outlier rate are both very
high and the photo-$z$'s are not very useful for any science applications.
The mean bias is small ($+0.01$), but this is because a negative bias
at $z<1$ is compensated by a positive bias at higher redshift by chance.
The biggest failure mode is that a large fraction of low-$z$ galaxies
up-scatter to high-$z$.  This is because the $griz$ photometry probes redwards of
the 4000\AA\ break of these galaxies, where there is no prominent spectral
feature.  The blue continuum of low-$z$ star forming galaxies in the optical
can be confused with blue UV continuum of galaxies at high redshifts,
resulting in the up-scattered population.  The overall poor photo-$z$
accuracy is the primary reason why SPS templates are not usually used for
photo-$z$'s and the two-step procedure with different templates is often
adopted to estimate, e.g., stellar mass.
We focus on a limited number of filters here, but SPS templates still
give poorer accuracy than empirical templates even when more filters are used.
E.g., \citet{dahlen13} carried out photo-$z$ tests with 14 filters for a number of codes and
the typical performance of SPS-based codes is worse than that of empirical codes.

If we apply the physical priors (case 2 in Table \ref{tab:photoz}
and Fig. \ref{fig:cosmos_photoz}), both the scatter and outlier
rate significantly decrease.   In particular, our primary failure mode
of the up-scattered low-$z$ galaxies is suppressed.
This is an important
result and it proves the concept of our technique;
the physical priors are indeed useful for improving photo-$z$'s.
Public photo-$z$ codes use a fixed set of SED templates regardless
of redshift.  The physical priors employed here effectively
evolve the templates with redshift and the observed improvements suggest
that the template evolution is actually important for photo-$z$'s.
We have not used information like sizes and morphologies here and
there is clearly room for further improvements.

We then apply the standard $N(z)$ prior (case 3).
As expected, this prior is efficient in removing low-$z$ galaxies
scattered out to high redshifts due to the degeneracy in
the multi-color space discussed above.
Both the overall scatter and outlier rate are reduced by the prior.
But, the backside of it is that it is also effective in reducing
correct photo-$z$'s at $z\gtrsim3$.
This is because those high redshift
galaxies are rare and $N(z)$ gives correspondingly small probabilities
at high redshift.  A preliminary analysis shows that this can
be remedied by applying stronger physical priors with a less
strong $N(z)$ prior (e.g., one can introduce a 'floor' in $N(z)$
such that the minimum probability at a given redshift is some
small fraction of the peak probability).
This is also an area where further work is needed to achieve
good photo-$z$'s across the entire redshift range.
We also should note that our $N(z)$ prior is constructed from
the \citet{ilbert09} catalog and we apply it to the same catalog here.
We will need to extensively test the prior on a different catalog
(see next section for comparisons with CFHTLenS).

In addition to the priors, we apply the template error functions (case 4).
The outlier rate ($f_{outlier}$) slightly increases, but this is
due to the reduced $\sigma$ (recall that outliers are defined as $>2\sigma$).
In fact, $f_{outlier,conv}$, which is based on the fixed outlier definition,
is reduced.  The overall accuracy is improved by the template error functions
and this means that the systematic mismatches between the SPS templates
and observed SEDs are in fact an issue in photo-$z$ and they have to
be corrected.  The implication here is not only for
SPS models but for observed SED templates as well;
a care must be taken when one applies observed SED templates collected at
$z=0$ to high redshift galaxies because there will be systematic
mismatches in the SEDs.

One may notice that there is a horizontal feature at $z_{phot}\sim0.5$
both at low-$z$ and high-$z$.
Only faint ($i>24$) galaxies contribute to this feature.
These faint galaxies have a wide and relatively flat PDF at
$0<z\lesssim0.8$, due to the large photometric uncertainties and
to the limited filter set.
For such PDF, the median redshift tends to cluster around $z_{phot}\sim0.5$.
At higher redshift, more than 2 filters fall below the 4000\AA\ break
and we can achieve a reasonable accuracy until we get to $z\sim3$,
where the $g$-band starts to fall below the Lyman break and again
the photometric uncertainty becomes very large.  
As shown in the next section, the additional $u$-band largely removes
this feature.

Our overall photo-$z$ accuracy may not appear as good as those achieved in,
e.g., DES \citep{sanchez14}.  But, there are important differences 
in the depths, the redshift range considered, and analysis method used.
\citet{sanchez14} studied galaxies with $i<24$ at $z<1.5$
and clipped 10\% of the objects with large uncertainties, while
we look at those with $i<25$ out to $z\sim5$ without any clipping.
We will make a fair comparison with an external code in Section 7.
Note that, for weak-lensing applications, the photo-$z$'s in 
Fig. \ref{fig:cosmos_photoz} are clearly not good enough and we do
need to apply a clipping technique such as the one developed by
\citet{nishizawa10} in order to reduce outliers.

%----------------
\subsection{Bias}

\begin{figure}[h]
\epsscale{1}
\plotone{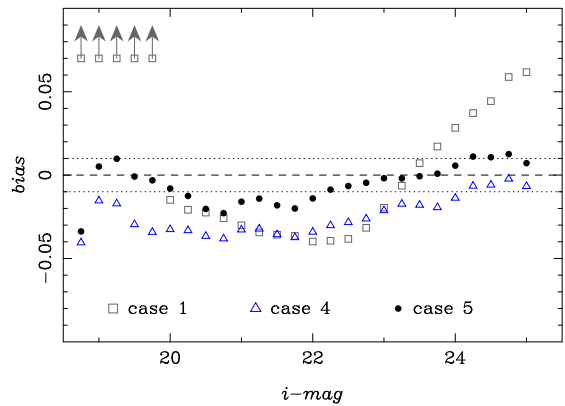}
\caption{
  Bias as a function of $i$-band magnitude.  The symbols are for
  different cases as shown in the figure.  Case 2, 3 and 5 are omitted
  for the sake of clarity.  The dashed line is $bias=0$ and the dotted
  lines are $bias=\pm1\%$.
}
\label{fig:bias_refmag}
\end{figure}

There is an issue with our photo-$z$ case 4, which is that
the photo-$z$'s are biased low, bias=$-0.019$.  This is illustrated
in Fig. \ref{fig:bias_refmag}, in which the bias is plotted as
a function of $i$-band magnitude.
This bias is a serious
problem for weak-lensing cosmology, which requires mean redshift
to be more accurate than 1 percent (e.g., \citealt{shirasaki14}).

Interestingly, this systematic offset has already been observed by
\citet{dahlen13}, who analyzed deep HST data with several photo-$z$
codes.  As shown in their Fig. 4, it is quite striking that most of
the codes (even for the same code with different setups) show the same
systematic offset of $\sim-0.01$.  Interestingly, the photo-$z$
comparisons by \citet{sanchez14} seem to suggest that machine-learning
techniques do not show such an offset.
It may be that a negative bias is a common problem in the SED-based
photo-$z$ techniques.
We find that the amount of the offset is dependent on the available data;
if we use all the 30-band photometry in COSMOS, the bias is essentially
zero (in fact, \citealt{ilbert09} observed no bias), but the bias
increases with decreasing the number of filters.

The negative bias has been observed but its origin has not been
discussed in the literature.  We have carefully looked into this issue,
but we have not identified the root cause of the bias yet.  We find that
the $N(z)$ prior introduces a negative bias, but it does not give
a full account of the observed bias.
We also find that the physical priors do not introduce a strong
bias (see case 6 discussed below).  The fact that public photo-$z$
codes that do not use physical priors also show a negative bias
supports this finding.
Just for now, we employ an ad-hoc correction to reduce the bias.
Because we know we tend to underestimate redshifts by 0.02,
we construct the template error functions with redshifts fixed
to $z_{spec}+0.02/(1+z_{spec})$.
The error functions constructed at offset redshifts reduce
the bias as shown in Table \ref{tab:photoz}, Figs. \ref{fig:cosmos_photoz}
and \ref{fig:bias_refmag} (case 5), although they slightly
increase the outlier rate.
The bias is not completely gone as shown in Fig. \ref{fig:bias_refmag}
and further calibrations may be needed.  We emphasize again that
the negative bias is not a specific problem to our code, but a common problem.

Finally, as the last proof of the effects of the physical priors, 
we show that photo-$z$'s still improve even when they are applied after
all the other corrections.  Case 6 in Fig. \ref{fig:cosmos_photoz}
and Table  \ref{tab:photoz} shows photo-$z'$s computed without
the physical priors.   A comparison with case 5   shows that
the physical priors are indeed effective in reducing the outliers
and dispersion even when they are added at the end.
Based on all the above results, we conclude that the physical priors improve photo-$z$'s.
As a further illustration of the physical priors, we show how photo-$z$'s change
if we change the prior parameters in the Appendix.

%---------------------
%---------------------

%----------------
\subsection{Accuracy of P(z)}

An important quality control of photo-$z$ is to assess the accuracy of PDF.
It is difficult to estimate how accurate individual PDFs are, but
the PDF accuracy can be addressed statistically.  We compute fractions of
objects that have photo-$z$'s consistent with the 30-band photo-$z$'s within
68 and 95 percentile intervals.  If the PDFs are perfect, the fractions should
be 68 and 95\%.  Deviations from these numbers are a manifestation of
incorrect PDF.

In case 5, we find that 61\% and 92\% of objects have consistent photo-$z$'s
within 68 and 95 percentile intervals, respectively.
It appears that our uncertainty estimates are slightly underestimated. 
\citet{dahlen13} observed a similar trend for a number of codes (see their Table 5).
This seems to be another common problem in SED-based photo-$z$ codes.
Catastrophic outliers probably contribute to the problem here and
the reduction of those outliers will help.  A careful construction
of the template error functions will be useful too (recall that
these functions include template uncertainties and thus are able to
control photo-$z$ uncertainties).  But, we do not perform further
calibrations at this point because $P(z)$ is strongly dependent on
the accuracy of photometric uncertainties, which is not always
straightforward to precisely measure.  The fact that \citet{ilbert09}
increased the flux uncertainty in the catalog by a factor of 1.5
does not seem to imply that the uncertainties are precisely measured.
We also find that reduced $\chi^2$ of bright objects is often too large
if we do not add a 0.02~mag uncertainty in the quadrature.
We defer further work on $P(z)$ to our future paper.

%----------------
\subsection{Brief summary}

We have demonstrated that the physical priors and template error functions
improve photo-$z$'s.  There are still issues left (e.g., understanding
the root cause of the negative bias) and there is still room for improvements
(e.g., priors on size and morphology), but the concept of the physical priors
is proven to be useful.  This is an important result.
We are now motivated to compare our code with an external code to
characterize how well our code works in the following sections.

%-------------------------------------------------------------------
%-------------------------------------------------------------------
\section{Comparison with CFHTLenS}

In this section, we compare our code with an external, publicly available code.
We do not run an external code by ourselves in order to
avoid any human biases (e.g., one may not fully calibrate an external
code to make his/her own code work better).  Instead, 
we use those from CFHTLenS \citep{heymans12},
which have been carefully computed by \citet{hildebrandt12}
with one of the most popular public codes {\sc bpz} \citep{benitez00},
to make a fair comparison.

%---------------------
\subsection{Data}

The photometric redshift in CFHTLenS is based on data from the CFHT Legacy
Survey.  The $ugriz$ photometric data are obtained with MegaCam under good
seeing conditions.  We have retrieved the photometric data from the CFHTLenS
database \citep{erben13}.  CFHTLenS does not overlap with COSMOS and we here
use spectroscopic redshifts from VVDS, not high-quality photo-$z$'s as done
in the previous section.  We focus on two VVDS layers: VVDS-Deep and VVDS-UltraDeep,
which are flux-limited to $i=24$ and to $23<i<24.75$, respectively \citep{lefevre13}.
The surveys are performed with VIMOS on VLT with different instrumental setups,
resulting in different redshift coverages.  We  discuss them separately.
We select secure spectroscopic redshifts with flags 3 and 4 from
these surveys and then cross-match them with the photometric objects from
CFHTLenS within 1 arcsec.  In total, we have 3,522 and 450 objects for VVDS-Deep 
and UltraDeep, respectively.

%---------------------
\subsection{Comparison with {\sc bpz}}

\begin{figure*}
\epsscale{1}
\plottwo{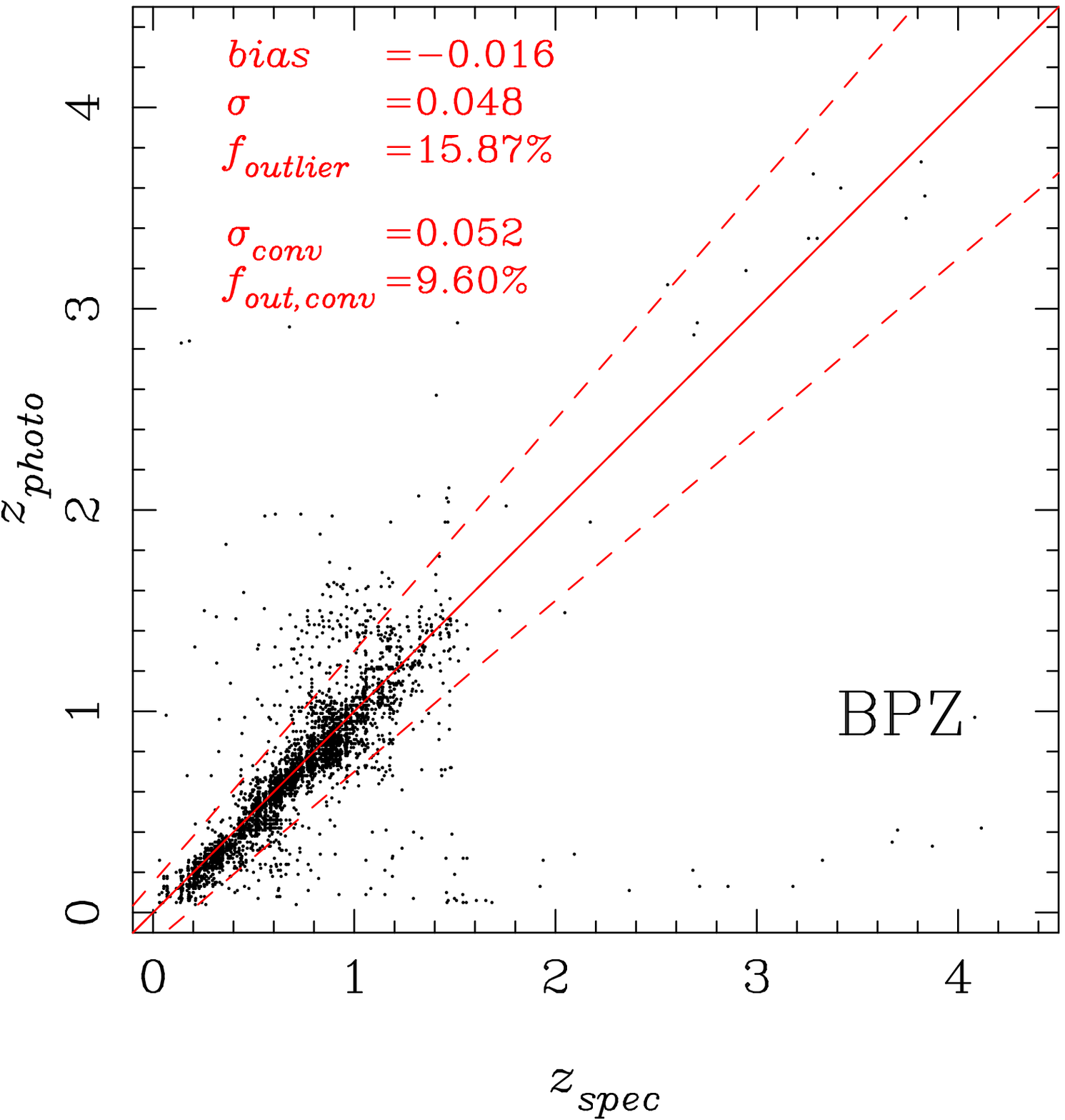}{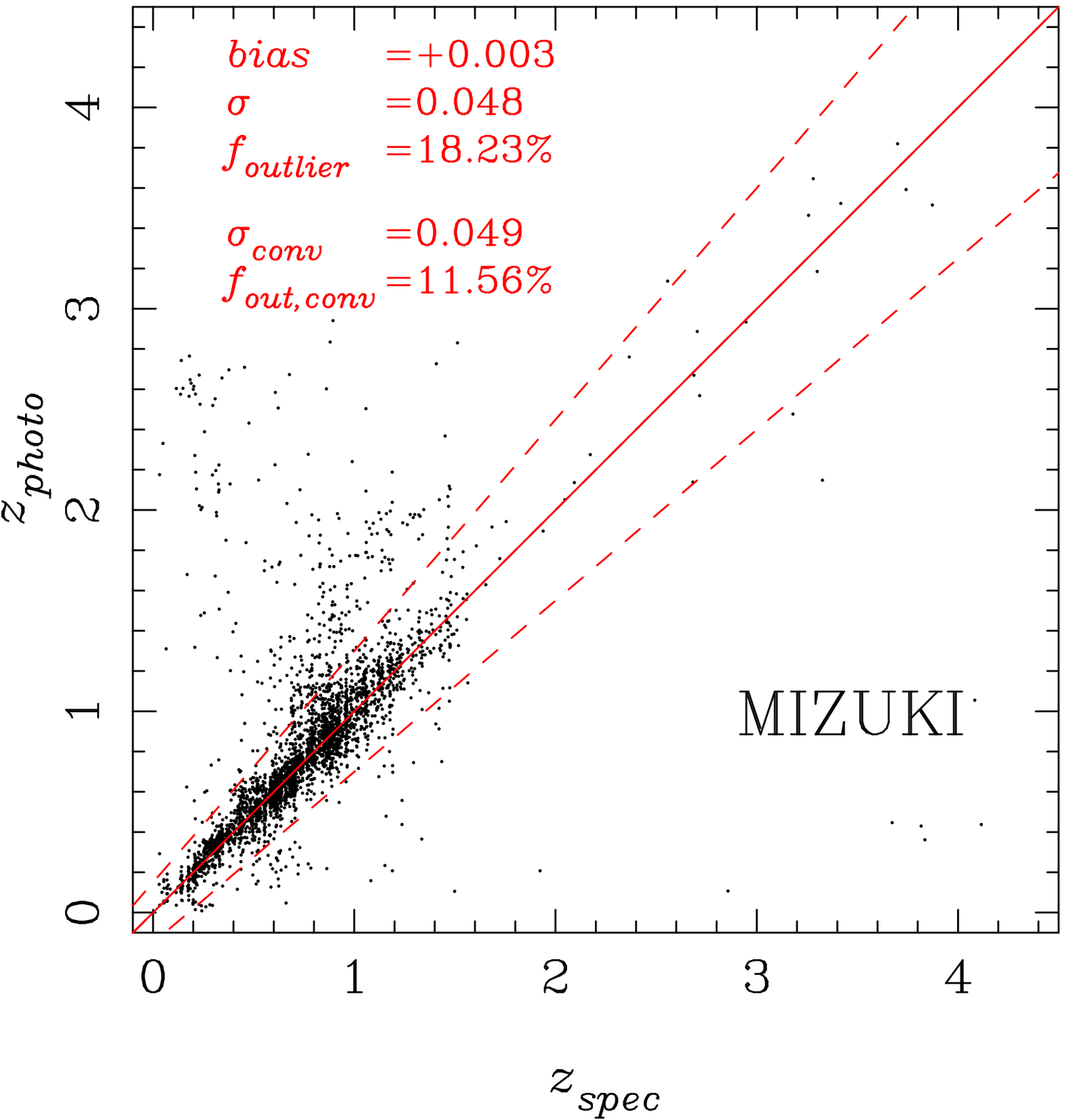}
\caption{
  $z_{phot}$ plotted against $z_{spec}$ from VVDS-Deep.  The panels are
  for {\sc bpz} (left) and {\sc mizuki} (right).  The solid and dashed lines
  show $z_{phot}=z_{spec}$ and $(z_{phot}-z_{spec})/(1+z_{spec})\pm0.15$, respectively.
}
\label{fig:vvds_d}
\end{figure*}

Let us start with VVDS-Deep shown in Fig. \ref{fig:vvds_d}.
The two codes perform similarly well; {\sc mizuki} shows
a higher outlier rate but with a slightly smaller scatter.
The failure mode of {\sc mizuki} is that a fraction of low-$z$ objects
up-scatter to high-$z$.  Such failures are fewer in {\sc bpz}.
An interesting point here is that {\sc bpz} has a small negative
bias of $-0.016$.
This negative bias again might be a common issue for SED fitting
(see discussion in the previous section).
On the other hand, {\sc mizuki}'s bias is only +0.003 thanks to
the template error function.

\begin{figure*}
\epsscale{1}
\plottwo{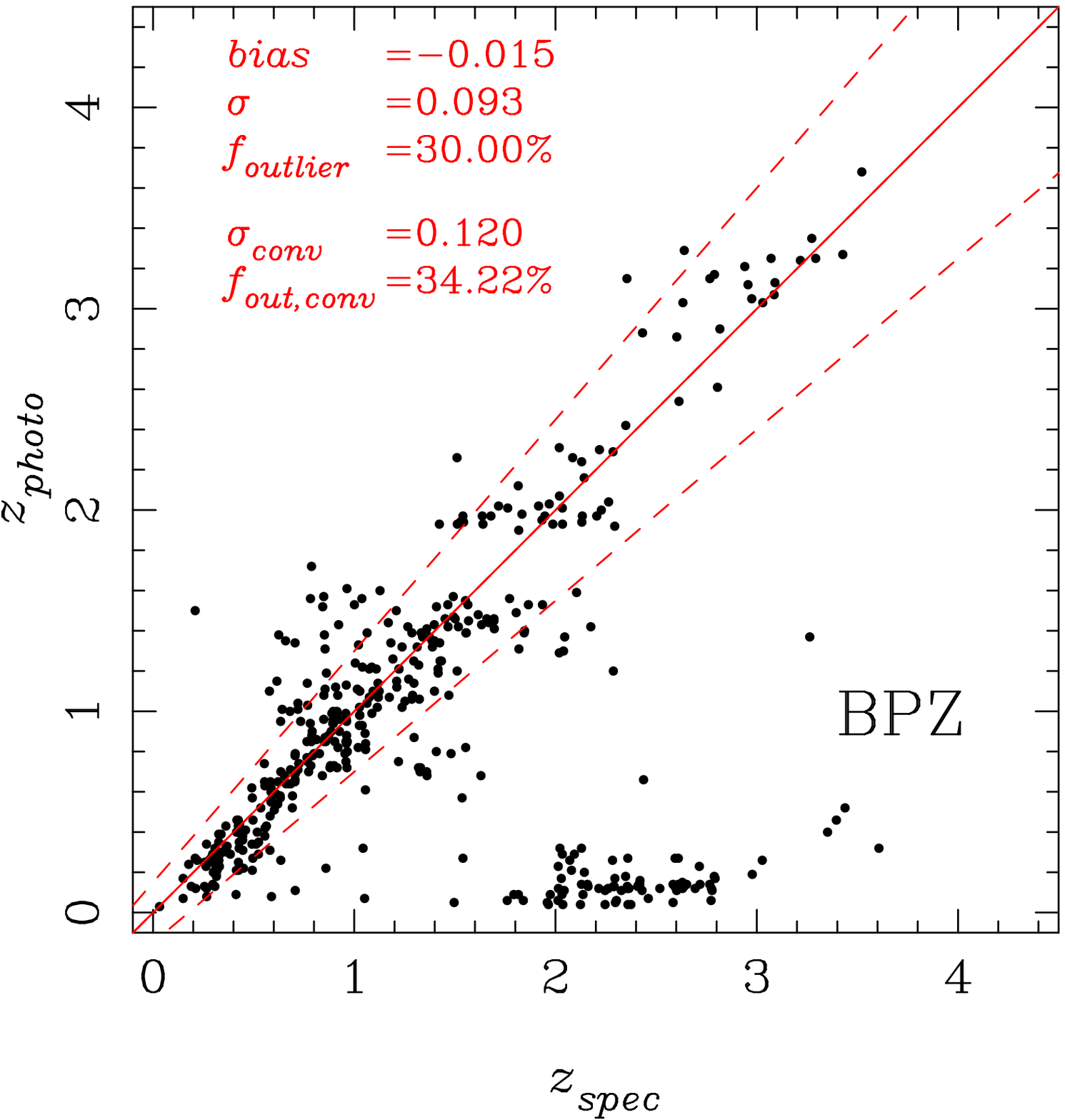}{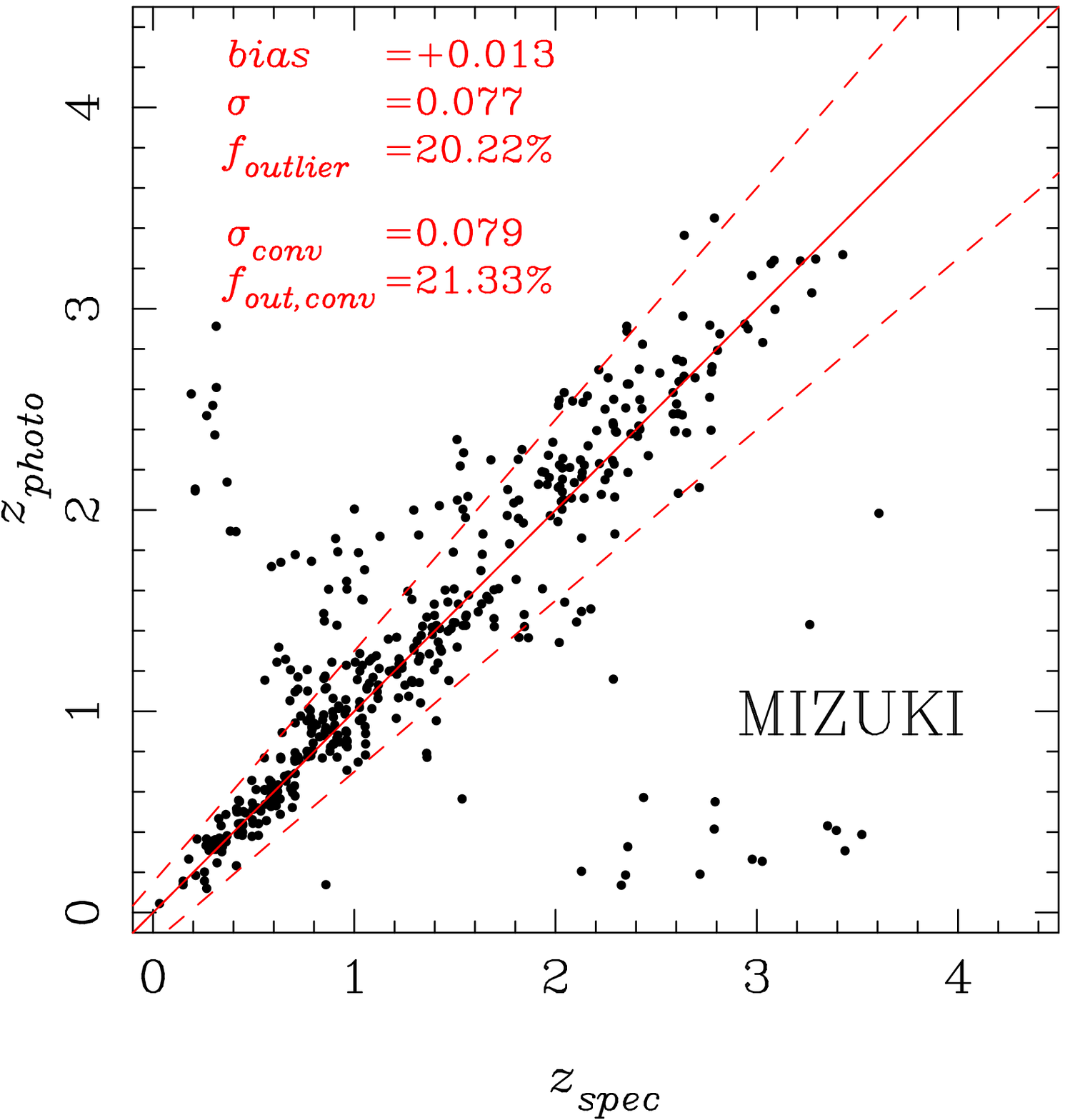}
\caption{
  As in Fig. \ref{fig:vvds_d}, but for VVDS-UltraDeep.
}
\label{fig:vvds_ud}
\end{figure*}

Turning to VVDS-UltraDeep in Fig. \ref{fig:vvds_ud}, there are
objects all the way out to $z\sim3.5$.  Although the instrumental
redshift selection function is probably not completely uniform,
VVDS-UltraDeep is less biased compared to VVDS-Deep thanks to
the very long integration time and a wider wavelength coverage.
We find that {\sc bpz} tends to give very low $z_{phot}$ to galaxies at $2<z_{spec}<3$,
which likely comes from the confusion between the Lyman break and
the 4000\AA\ break.  It might be that the $N(z)$ priors adopted
in the code strongly prefer the low-$z$ solution, resulting
in the down-scattered population.
On the other hand, {\sc mizuki} does not show such a strong failure
and the numbers of outliers at low-$z$ and high-$z$ are about
the same, suggesting the priors are  about right.

In summary, {\sc mizuki} is at least as good as {\sc bpz} and possibly
better at faint magnitudes and at high redshifts.
The physical priors work well and the template error functions
keep the bias smaller than that of {\sc bpz}.
While further calibrations may be needed, {\sc mizuki} is already
a powerful tool to infer photometric redshifts.

%---------------------
%---------------------
%---------------------
%---------------------

%-------------------------------------------------------------------
%-------------------------------------------------------------------
\section{Biases in physical properties of galaxies}

Physical properties and redshifts are very often
inferred using different templates in the literature; one first uses
empirical templates, PCA-based templates, or numerical techniques to 
compute redshift because they deliver better redshifts than SPS templates.
Then, one switches to SPS templates to infer physical
properties at the best-fit photo-$z$.  This is not a self-consistent
way to infer them because different template sets may give different
$P(z)$ and the spectral shape and redshift are partially degenerate.
Our code delivers good photo-$z$ accuracy and we can compute physical
properties in a self-consistent manner.
As of this writing, the code computes $P(M_*)$, $P(SFR)$, $P(\tau_V)$
marginalized over all the other parameters.

We use the template error functions
to remove small systematics in the SPS templates.  Also, we apply
the priors on physical properties.  They may affect the inferred properties.
In this section, we quantify biases in the inferred properties and show that they are well
within the systematics of the SPS models that are inherent in any physical
property estimates based on the SED fitting.  In general, the recovery
accuracy of physical properties is dependent on redshift accuracy,
filter set (and depths), and details of the SED fitting technique.
We have demonstrated in the previous sections that the template error
functions and physical priors improve photo-$z$. 
Also, intensive work on the recovery accuracy
for a number of filter combination is already done by \citet{pforr12}.
For these reasons, we fix redshifts
and assume only 2 representative filter sets
in order to make useful comparisons focused on the fitting technique.

In the following, we make two comparisons: internal comparison and external
comparison.  The first one is to use our code with and without the template
error functions and physical priors to quantify differences in the inferred
properties.  The 2nd one is to compare with a public SED fitting code
with identical assumptions in the SPS models.
As mentioned earlier, one should use the full PDFs of the physical properties
for science analyses, but it is nevertheless useful to make point estimates.
We take the median of PDF as done for redshift (see Eq. 21).

%---------------------
\subsection{Data}

For the purpose of the section, we need a public multi-wavelength catalog
with physical property estimates.  NEWFIRM medium band survey
(NMBS; \citealt{vandokkum09}) is a good resource for this.
NMBS achieved one of the most accurate photometric redshifts to date
thanks to the large number of broad-band and medium-band photometry.
This also allows reliable measurements of
physical properties of galaxies.
NMBS presents two catalogs: one in COSMOS and the other in AEGIS.
We use the COSMOS catalog here because it includes medium-band data
in the optical and is better than the other one \citep{whitaker11}.
We take the $griz$ photometry and photo-$z$'s from the catalog
for the analysis in the following section.

%---------------------
\subsection{Internal comparison}

\begin{figure}
\epsscale{1}
\plotone{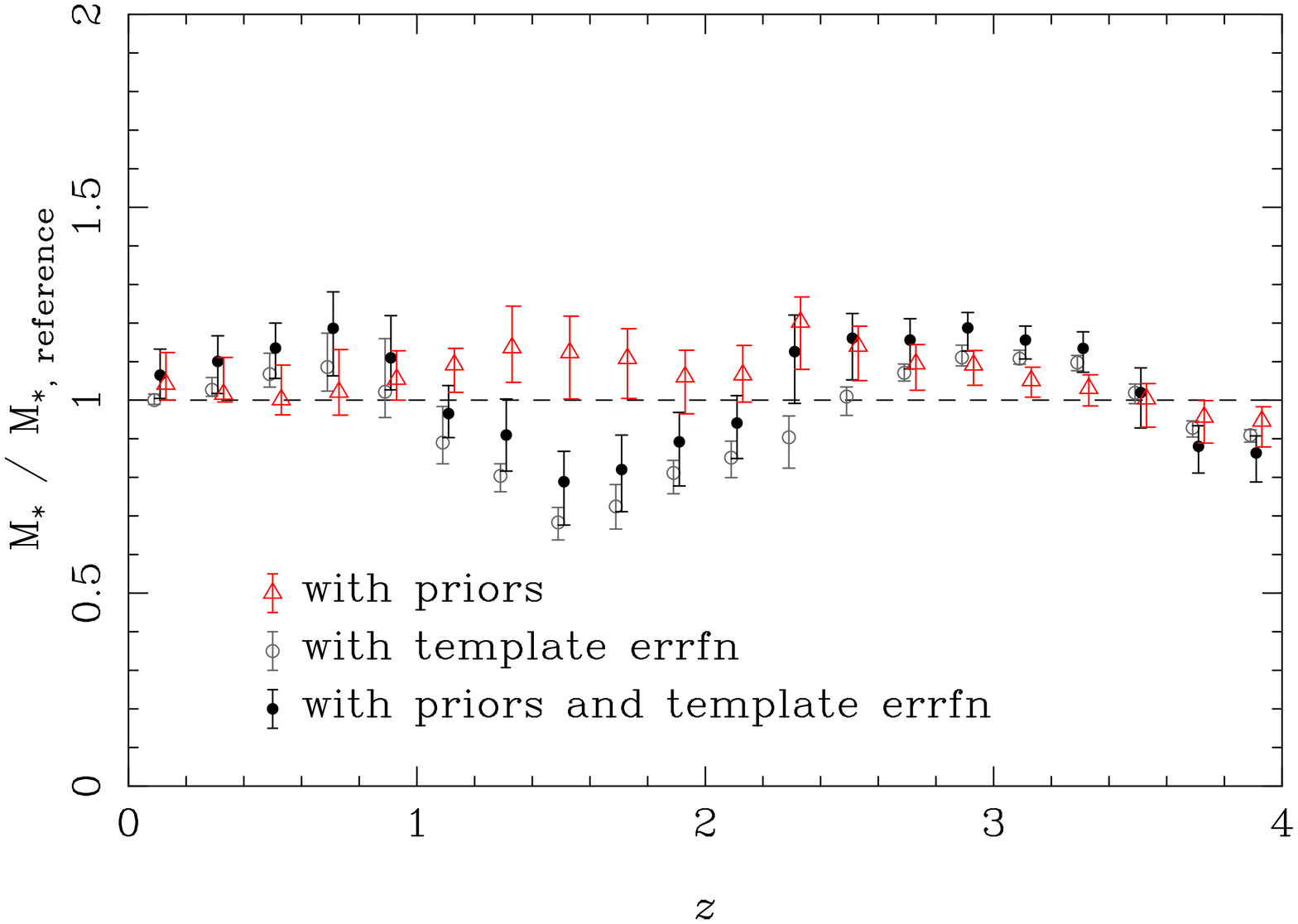}
\plotone{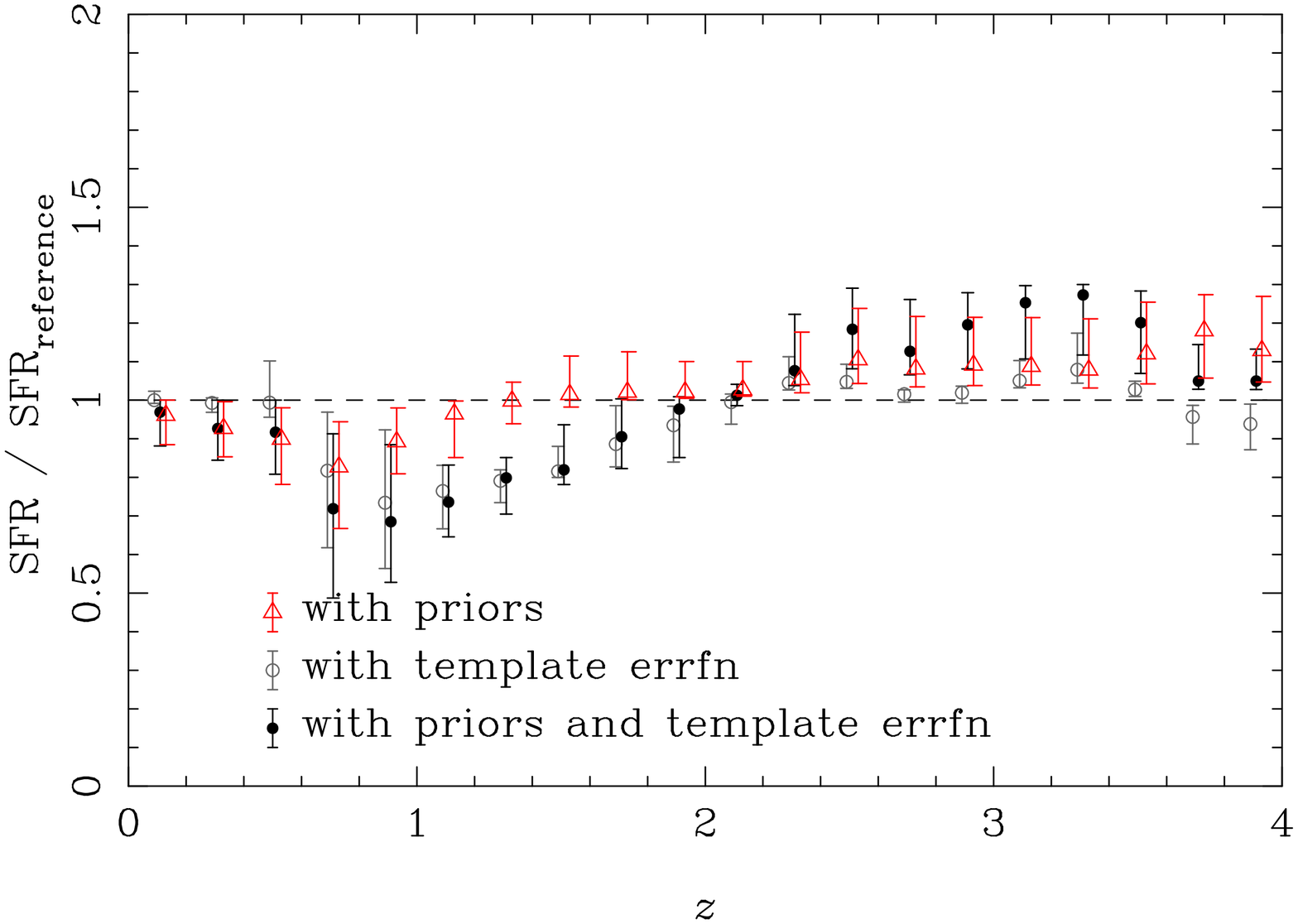}
\plotone{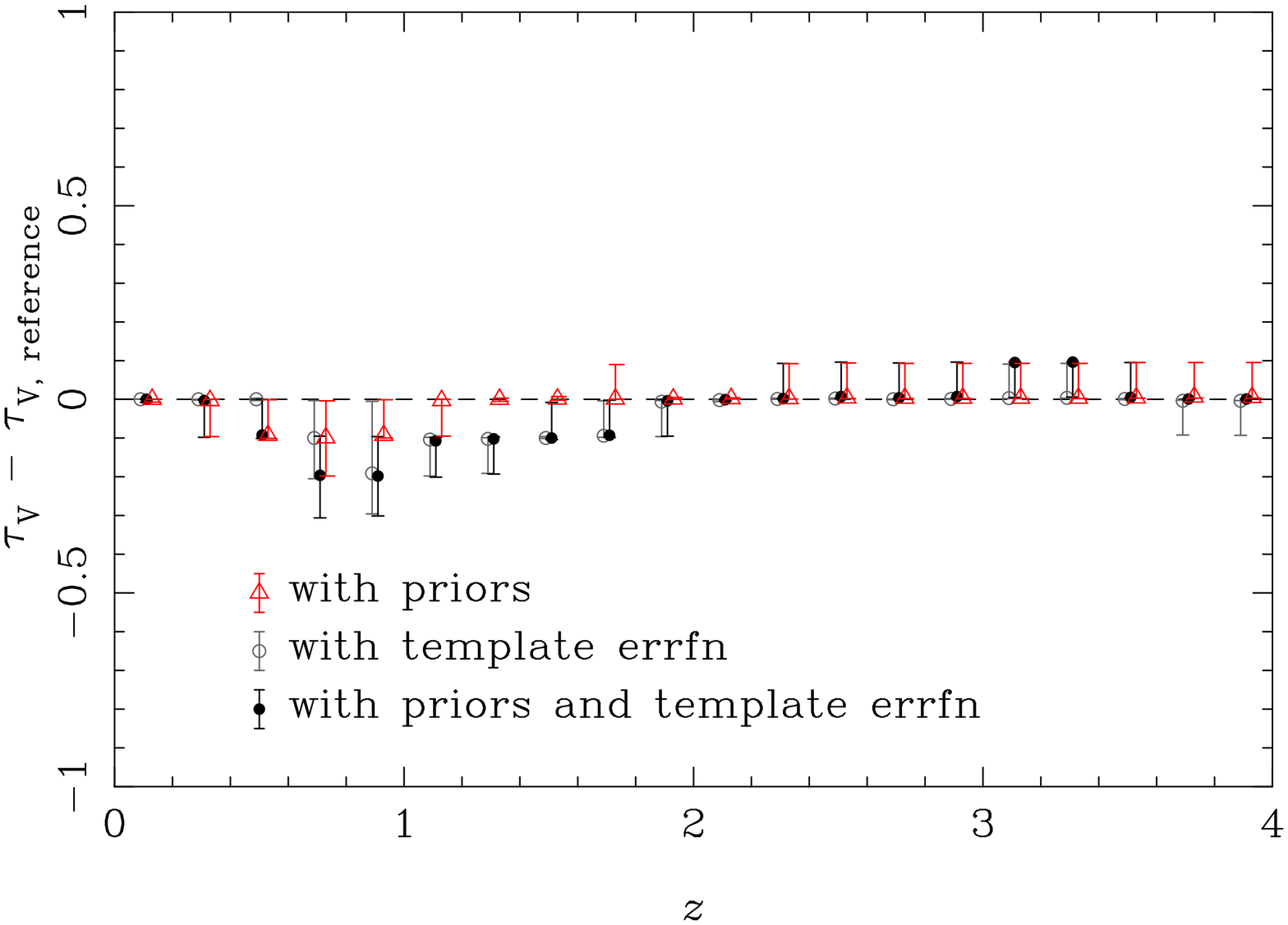}
\caption{
  Ratio or difference between physical parameters inferred with and without
  the template error functions and physical priors (legend shown in the plots).
  The points and the error bars show the median and quartiles of the distribution.
}
\label{fig:bias_phys}
\end{figure}

First, we make an internal comparison.  We compute stellar mass, SFR,
and $\tau_V$ with and without the template error functions and the physical priors.
The fits are performed at the NMBS photo-$z$'s.
Fig. \ref{fig:bias_phys} shows the differences in the inferred physical
properties as a function of redshift.
If only the priors are applied, there is only mild redshift dependence
with a weak bias up to $\sim20\%$.  The template error function introduces
stronger redshift dependence, giving rise to $\sim30\%$ bias at some redshifts.
The bias for the priors and error functions combined is similar to
that of the error function only case, which means that the overall bias is
primarily due to the template error function.
We know that SPS models do not perfectly
reproduce the observed SEDs of galaxies, even for passively evolving galaxies
\citep{maraston09}.  Most authors nevertheless fit SPS models and use the one
that is most similar to an observed SED to infer its physical properties.
The $\sim30$\% bias observed in Fig. \ref{fig:bias_phys} can be regarded as
a level of the systematic uncertainty in this process.
Fortunately, it is smaller than the overall systematic uncertainty
inherent in the physical properties; e.g., \citet{conroy09} suggested that
stellar mass measured at $z=0$ carries an uncertainty of about a factor of 2.
While the bias introduced by the physical priors and template error
functions should be kept in mind, it is unlikely to
dominate the overall error budget.

%---------------------
%---------------------

%--------------------
\subsection{External comparison}

%---------------------
\begin{figure*}
\epsscale{1}
\plottwo{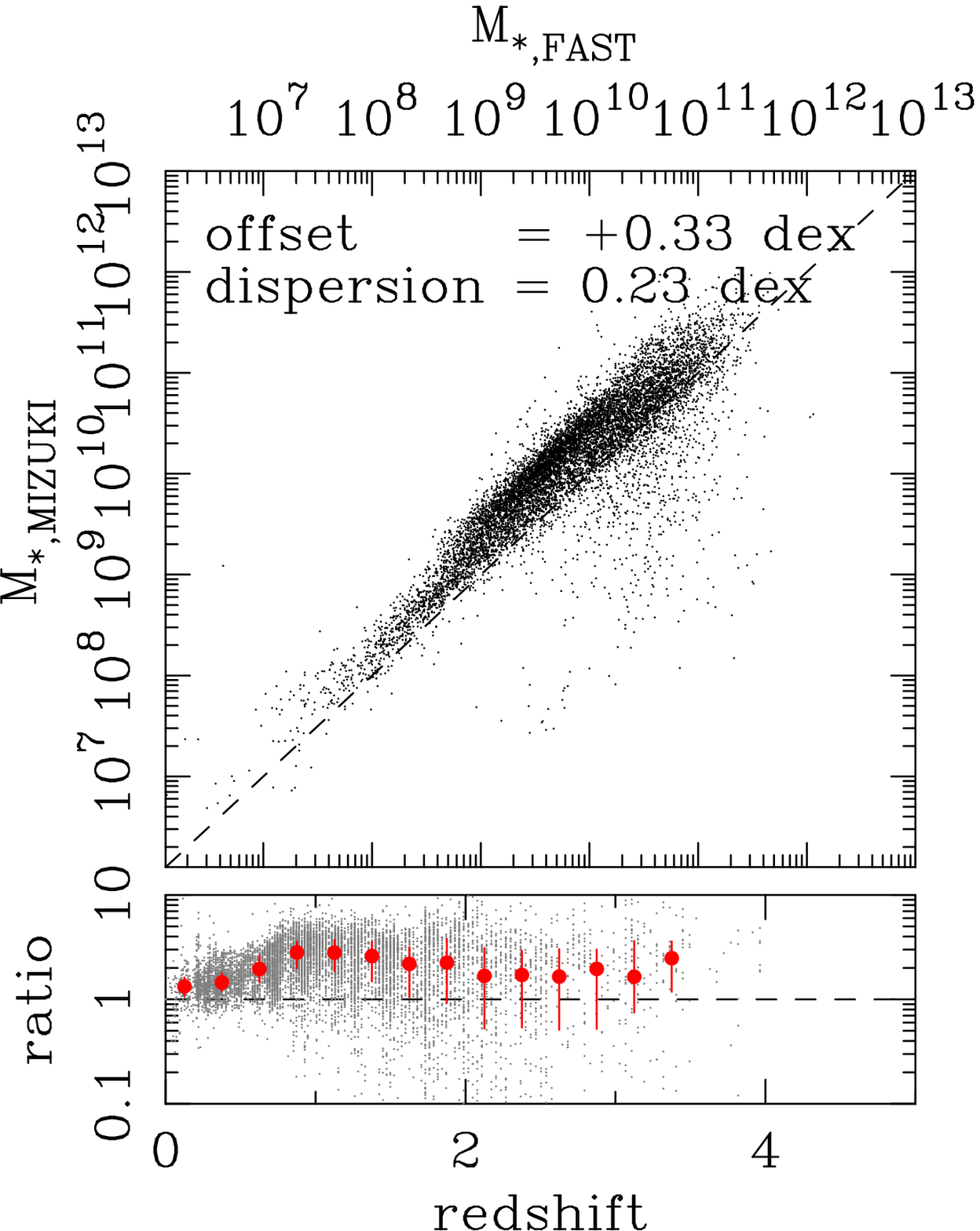}{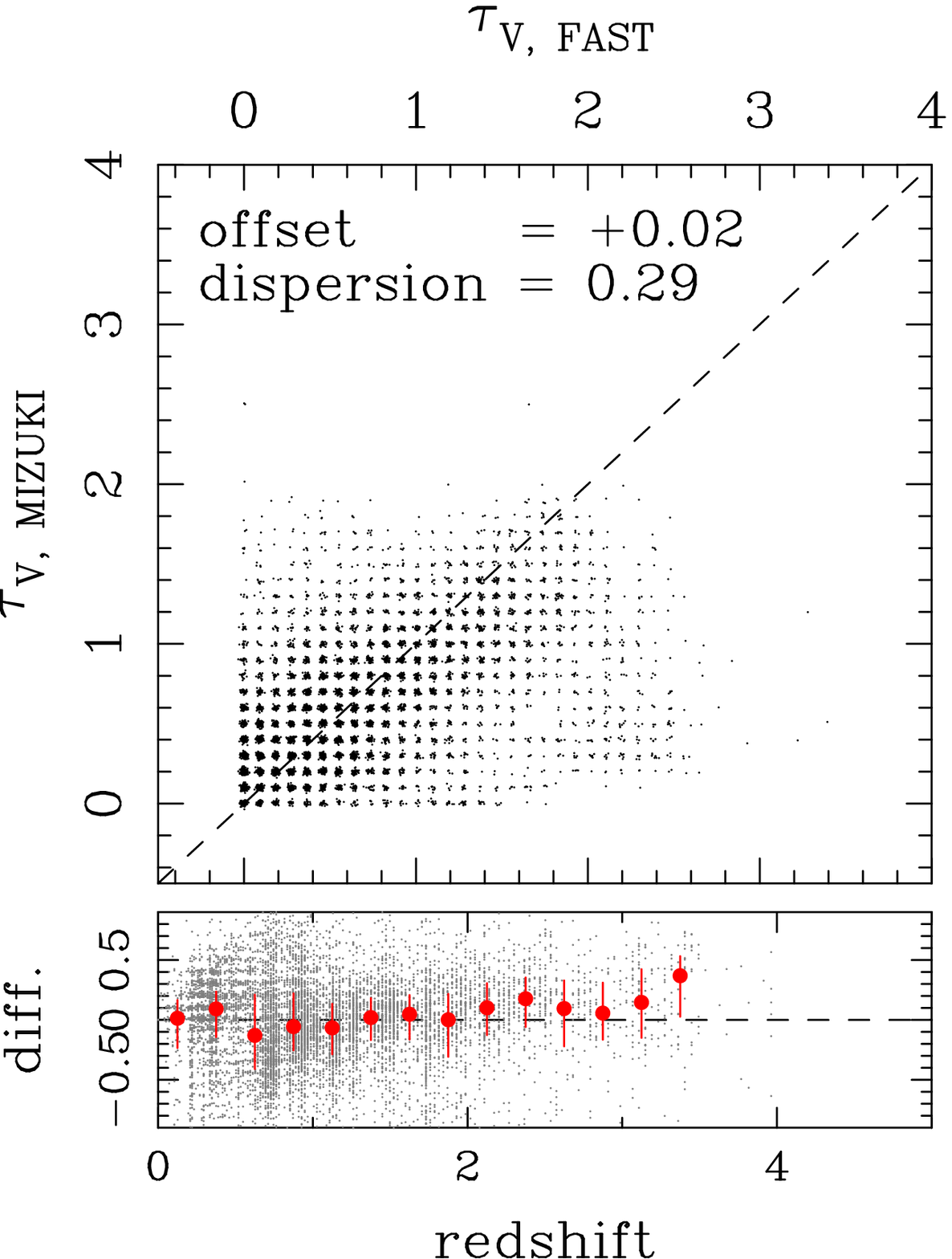}\vspace{0.5cm}
\plottwo{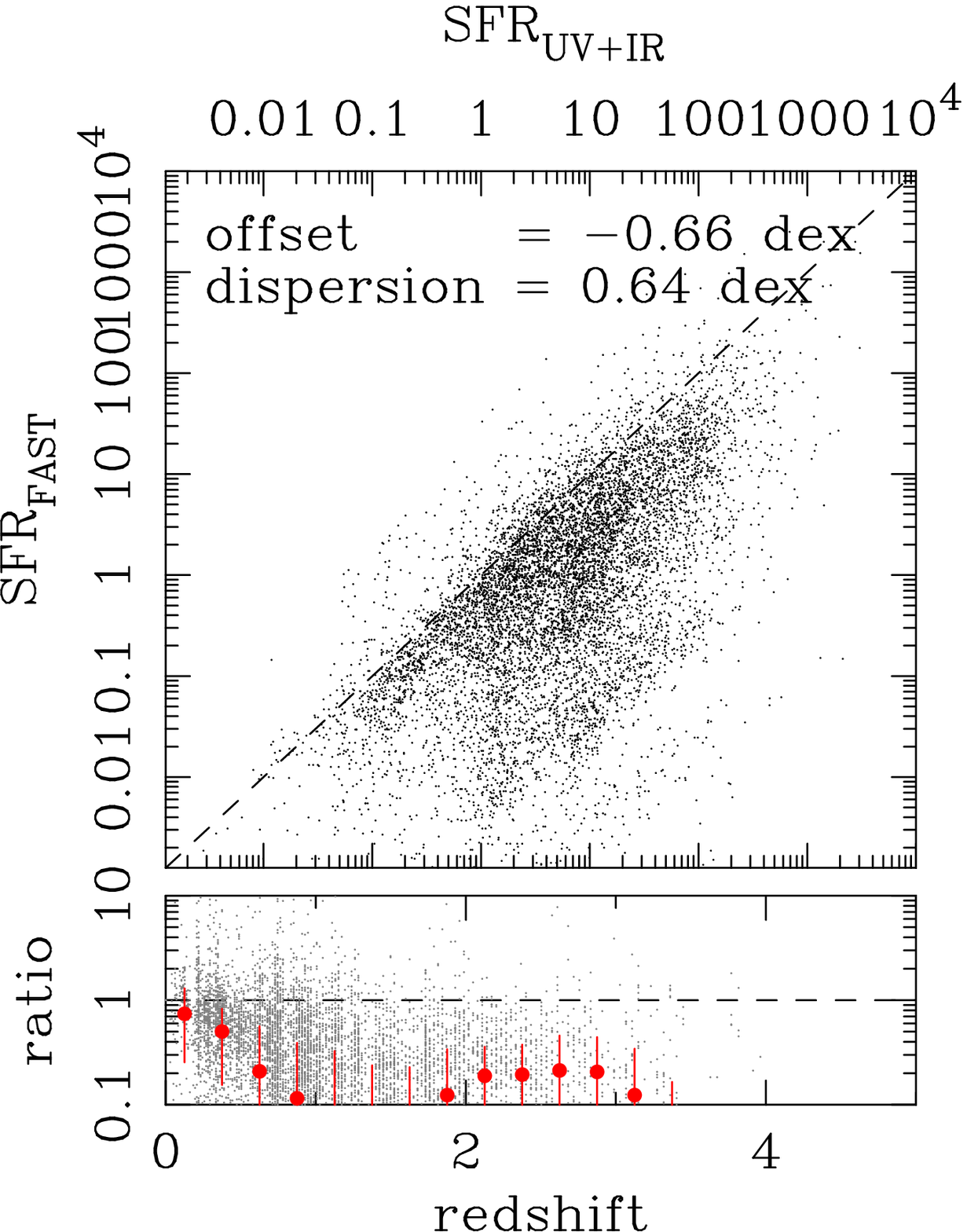}{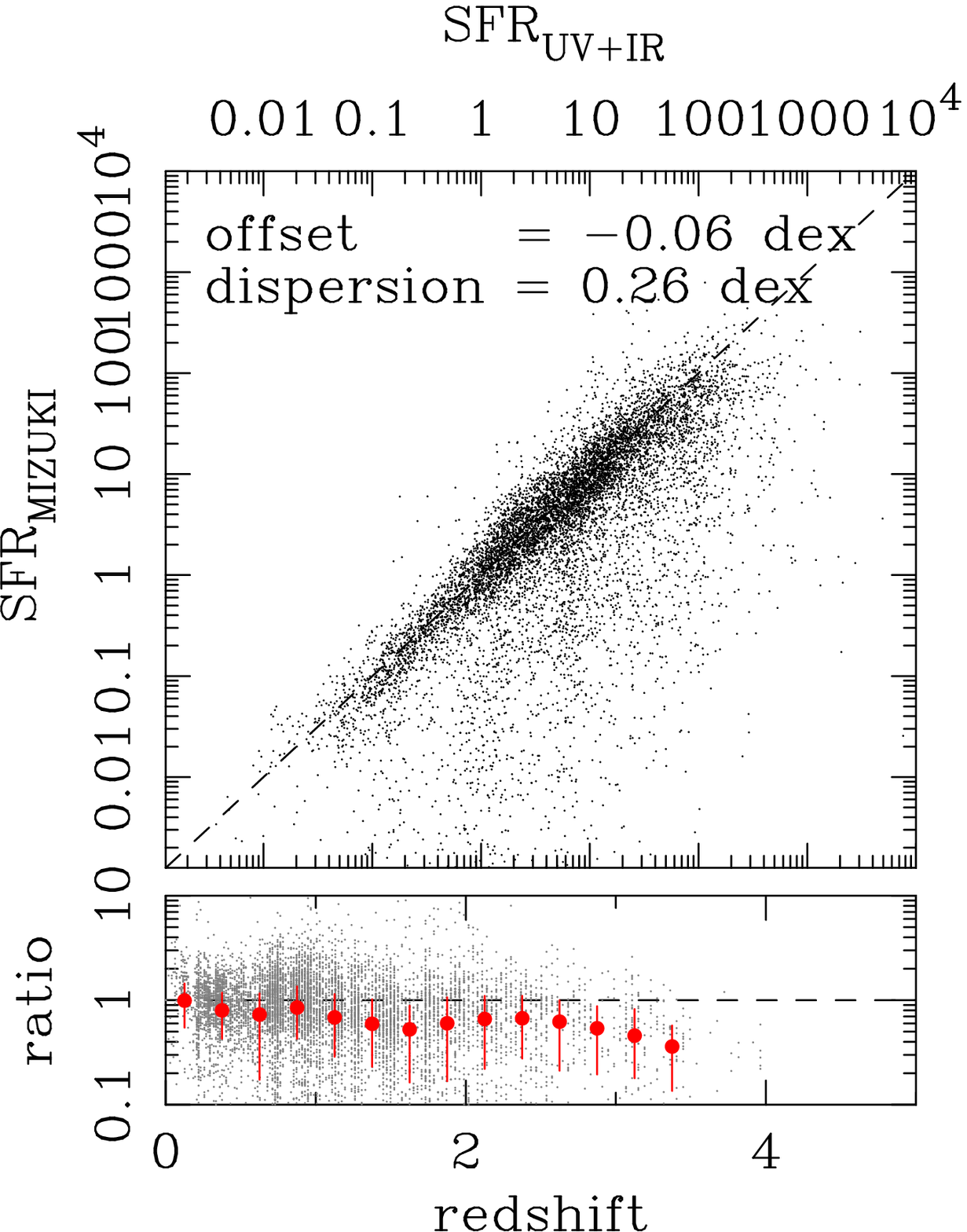}
\caption{
  Top plots show physical properties inferred with our code against
  those with {\sc fast}.  The top-left is for stellar mass and the top-right
  is for attenuation.   The bottom plots show correlations between
  different SFR estimators.  The bottom-left panel shows SFR$_{\rm FAST}$
  against SFR$_{\rm UV+IR}$ and the bottom-right panel is SFR$_{\rm MIZUKI}$
  against  SFR$_{\rm UV+IR}$. 
  The small panel in the bottom of each plot shows the ratio or difference between
  the vertical and horizontal axes as a function of redshift.
}
\label{fig:bias_phys2}
\end{figure*}
%---------------------

Next, we compare our code with {\sc fast} \citep{kriek09}.
{\sc fast} is a $\chi^2$ code to infer physical properties of galaxies by fitting
observed SEDs of galaxies with SPS models.
\citet{whitaker11} applied the code to the NMBS data under identical assumptions to ours:
\citet{bruzual03} SPS models with exponentially decaying star formation history, 
Chabrier IMF, Calzetti attenuation curve \citep{calzetti00}, and solar metallicity.
{\sc fast} is run with all the filters in the catalog, and a fair code-code comparison
can be made if we use all the filters.
But, we choose to primarily discuss the $griz$ photometry in what follows
in order to keep consistency with the previous sections and also to obtain
a crude sense of how accurately we can infer the physical properties using
optical data from large surveys.
We summarize all of our results in Fig. \ref{fig:bias_phys2} and
Table \ref{tab:recovery_accuracy}.
We discuss each physical property in turn.

{\bf Stellar mass:}
On average, our stellar mass is about a factor of 2 higher than {\sc fast}.
This difference may not be too surprising because we use only 4 filters,
while {\sc fast} used more than 35 filters over a wider wavelength range. 
At $z\gtrsim1$, all of the $griz$ filters fall below the 4000\AA\ break and
we lose an ability to constrain the stellar mass to luminosity ratio,
resulting in the large bias and scatter at high redshifts. 
As shown in Table \ref{tab:recovery_accuracy}, the overall offset reduces
to $\lesssim0.1$~dex when we use all the available filters.
Thus, the limited photometry, not the code or templates, is likely
a primary cause of the offset.
The typical statistical uncertainty in our stellar mass is
0.2-0.3~dex, even though we fix the redshift in the fits.

We note in passing that similar redshift dependence was observed
by \citet{vandokkum14} in their comparison between the UltraVISTA and 3D-HST data.
They used a large number of filters to
infer stellar mass using {\sc fast}, but they still observed a difference in stellar mass
between the two data sets with a somewhat strong redshift dependence of $-0.3z$~dex.  
This seems to imply that, not only codes and templates, but data matter a lot.

{\bf Dust attenuation:}  
We find that our estimates are consistent with {\sc fast}, although
the dispersion is large $\sigma(\tau_V)\sim0.3$.
As discussed in \citet{pacifici12}, we expect a large uncertainty in 
the inferred amount of attenuation from the broad-band photometry
and we should first note that the attenuation inferred by {\sc fast} is
not the truth table here.
With this caveat in mind, it is interesting that
we do not observe a significant systematic
difference at any redshifts.  This is somewhat surprising because 
our photometry probe only the UV wavelengths at high redshifts.
It is likely because the amount of attenuation is controlled well
by the physical prior.

{\bf SFR:}
\citet{whitaker11} computed SFRs from the sum of the UV and IR
luminosities under the reasonable assumption that absorbed UV photons
are re-radiated thermally in the IR.  The bottom-left panel of Fig. \ref{fig:bias_phys2}
compares $\rm SFR_{UV+IR}$ with SFRs from {\sc fast} based on the SED fitting.
Assuming $\rm UV+IR$ is a reasonable tracer of SFR, we find that
{\sc fast} tends to underestimate SFRs.  At $z>1$, SFRs from {\sc fast} are lower by
an order of magnitude.  In contrast, {\sc mizuki} estimates SFRs reasonably well as
shown in the bottom-right panel.  SFRs are slightly underestimated, but
the overall agreement with $\rm SFR_{UV+IR}$ is good and the dispersion
(which is computed by clipping outliers as done in Section 6) is relatively small, a factor of $<2$.
Interestingly, the offset increases if we include all the filters
out to IR wavelengths (Table  \ref{tab:recovery_accuracy}).  One of the reasons
for this would be our templates do not include thermal emission from warm dust yet.

\vspace{0.5cm}
Overall, physical properties inferred by {\sc mizuki} agrees reasonably
well with {\sc fast}.  There is a redshift-dependent offset in stellar mass,
but it is likely due to the limited photometry.
Amount of attenuation agrees well, and for SFR, {\sc mizuki} works better
than {\sc fast}.
Together with the good photo-$z$ accuracy demonstrated in the previous section,
{\sc mizuki} is a powerful tool to infer physical properties
in a fully self-consistent manner.

%---------------------------------
\begin{table*}[ht]
  \begin{center}
    \begin{tabular}{ccccc}
      parameter                      & offset ($griz$)  & dispersion ($griz$) & offset (all) & dispersion (all)\\\hline
      $\rm M_{*,MIZUKI}/M_{*,FAST}$       & $+0.33$~dex     & $0.23$~dex          & $+0.09$~dex  & $0.12$~dex\\
      $\rm \tau_{V,MIZUKI}/\tau_{V,FAST}$ & $+0.02$         & $0.29$              & $-0.16$      & $0.25$\\
      $\rm SFR_{FAST}/SFR_{UV+IR}$       & ---             & ---                 & $-0.66$~dex  & $0.64$~dex\\
      $\rm SFR_{MIZUKI}/SFR_{UV+IR}$     & $-0.06$~dex     & $0.26$~dex          & $-0.34$~dex  & $0.25$~dex\\ 
      \hline
    \end{tabular}
  \end{center}
  \caption{
    Consistency between {\sc mizuki} and {\sc fast}.
  }
  \label{tab:recovery_accuracy}
\end{table*}
%---------------------------------

%-------------------------------------------------------------------
%-------------------------------------------------------------------
\section{Room for improvements}

Throughout the paper, we have identified areas where
further work is needed and there is clearly a large room for
improvements in our technique.
One of the ways forward would be to explore new priors.
We have assumed solar metallicity in the SSP models,
but given the recent substantial progress in the field, we could introduce
a mass-metallicity prior.  
The size and morphology information has not been fully exploited
in the literature (\citealt{wray08} may be a notable exception) and 
the size-mass and morphology-mass (or SFR) priors would be interesting
to explore.  Surface brightness might be useful, too \citep{stabenau08}.

We also need to improve our templates.  We have limited ourselves
to optical data in this paper, but multi-wavelength data sets are
often available in deep fields.  One urgent improvement would be to
include thermal emission from warm dust.  Because we know the amount of
attenuation and SFR for each template, we can add a model of
thermal emission to the stellar SED based on the energy conservation
(e.g., \citealt{dacunha08}).  In addition to galaxies, we are also
interested in stars and QSOs.  We are making progress in QSO and
stellar models as well as priors for them and we defer detailed
discussions on them to our future paper.

Finally, speed.  The current speed of the code is about 2 objects per second,
which is too slow to be applied to survey data.  The code can be optimized further
and the same is true for templates; we find that 20-30\% of
the templates generated in Section 3 never give good fits to the observed
SEDs of galaxies.  We can safely remove those templates from the library,
which will make the code faster.  We aim to address these issues in our
future work.

%-------------------------------------------------------------------
%-------------------------------------------------------------------
\section{Summary}

We have presented a proof-of-concept analysis of photometric redshifts
with Bayesian priors on physical properties of galaxies such as SFR,
stellar mass and dust attenuation.  The priors are not fully optimized yet,
but we have shown that the priors improve photometric redshifts
significantly.  Furthermore, template error functions, which are
intended to correct for systematic flux errors as well as to assign
random uncertainties to model templates, also improve photometric redshifts.
We have compared our code with {\sc bpz}, which is one of the most popular
photo-$z$ codes, and shown that {\sc mizuki} performs similarly well
at bright magnitudes and it works better at faint magnitudes and
at high redshifts.

One unique feature of the technique is that we can simultaneously
infer redshifts and physical properties of galaxies in a fully
self-consistent manner, unlike the two-step procedure with different
templates often adopted in the literature.
We have compared physical properties inferred by {\sc mizuki} with those
from {\sc fast} under identical assumptions and confirmed that the inferred
properties agree well, except for SFR, for which {\sc mizuki} works
significantly better.
The priors and template error functions inevitably introduce a bias
in the inferred physical properties of galaxies, but we have shown that
it is small, $\lesssim30\%$.  The bias is primarily due to mismatches
between SPS SEDs and observed SEDs of galaxies and hence it is
a problem common to all codes that
use the \citet{bruzual03} model to infer physical properties.

Overall, {\sc mizuki} is a powerful code to infer both redshifts and
physical properties of galaxies.  We have identified a number of
areas where further work is needed throughout the paper.  We will
improve the code using data from the on-going large surveys such as
the HSC survey.  Once the code becomes mature enough, we will
make the code available to the public.

%-------------------------------------------------------------------
\bigskip

We thank the HSC photo-$z$ working group, especially Atsushi Nishizawa
and Jean Coupon, for useful discussions on many aspects of photo-$z$,
and the anonymous referee for useful comments.
We acknowledge support by KAKENHI No. 23740144.

This work is based on observations obtained with MegaPrime/MegaCam,
a joint project of CFHT and CEA/IRFU, at the Canada-France-Hawaii
Telescope (CFHT) which is operated by the National Research Council
(NRC) of Canada, the Institut National des Sciences de l'Univers of
the Centre National de la Recherche Scientifique (CNRS) of France,
and the University of Hawaii. This research used the facilities of
the Canadian Astronomy Data Centre operated by the National Research
Council of Canada with the support of the Canadian Space Agency.
CFHTLenS data processing was made possible thanks to significant
computing support from the NSERC Research Tools and Instruments grant program.
This research uses data from the VIMOS VLT Deep Survey, obtained from
the VVDS database operated by Cesam, Laboratoire d'Astrophysique
de Marseille, France. 
This study makes use of data from the NEWFIRM Medium-Band Survey,
a multi-wavelength survey conducted with the NEWFIRM instrument
at the KPNO, supported in part by the NSF and NASA.

%-------------------------------------------------------------------------------
%-------------------------------------------------------------------------------
\appendix

%-------------------------------------------------------------------
\section{Effects of priors and template error functions on a $K$-selected catalog}

%---------------------
\begin{figure*}
\epsscale{1}
\plottwo{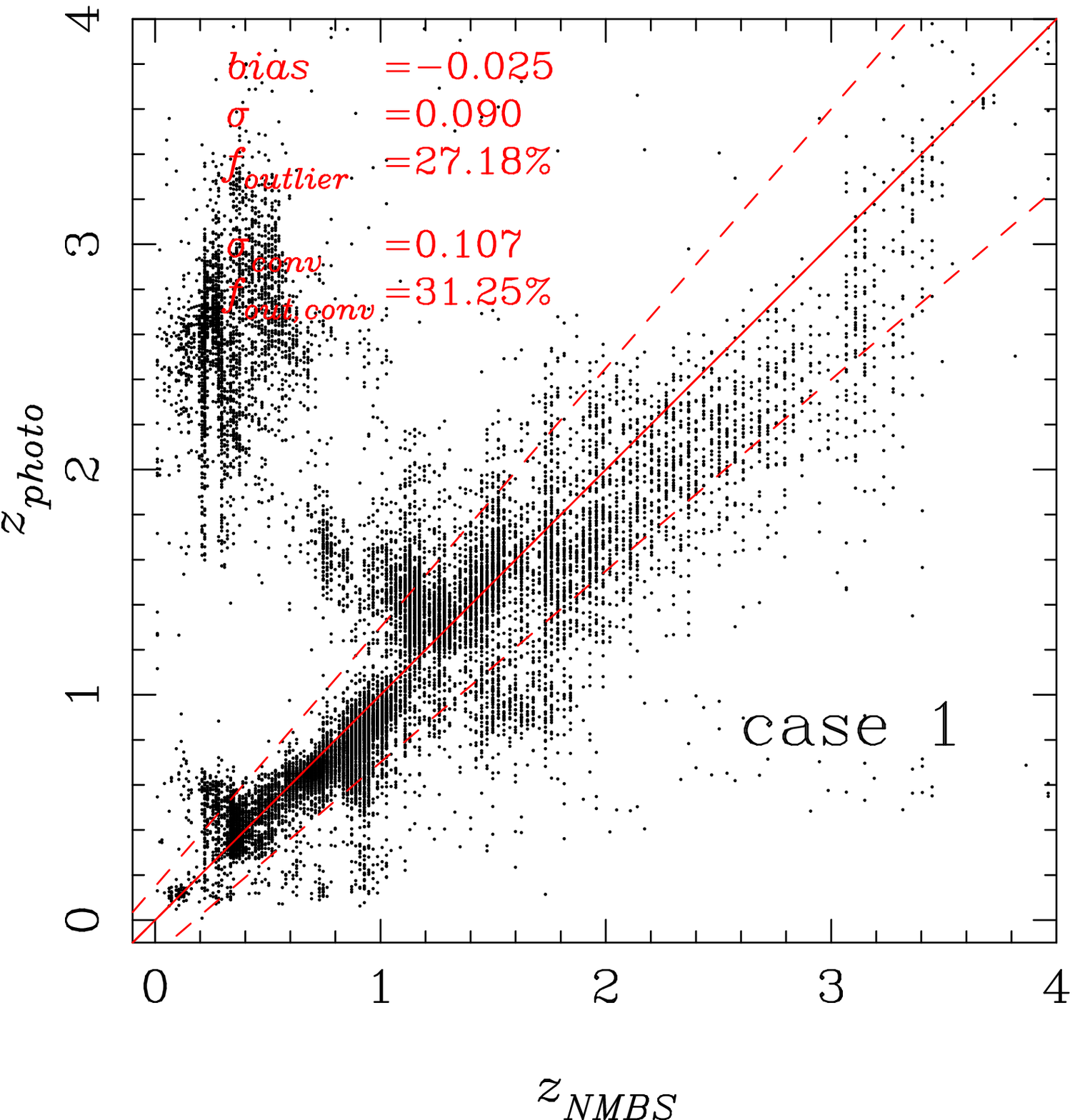}{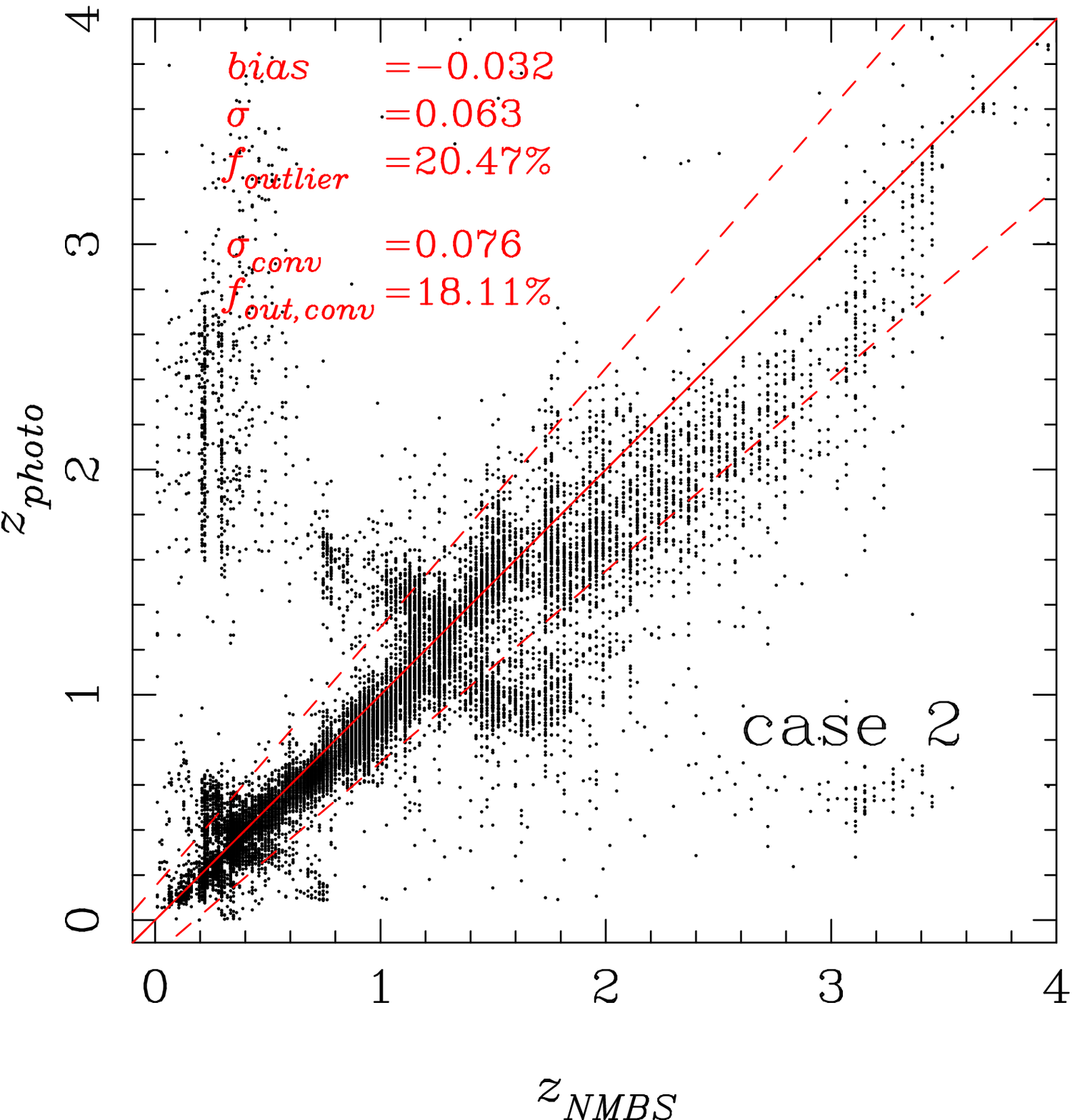}
\includegraphics[scale=0.5]{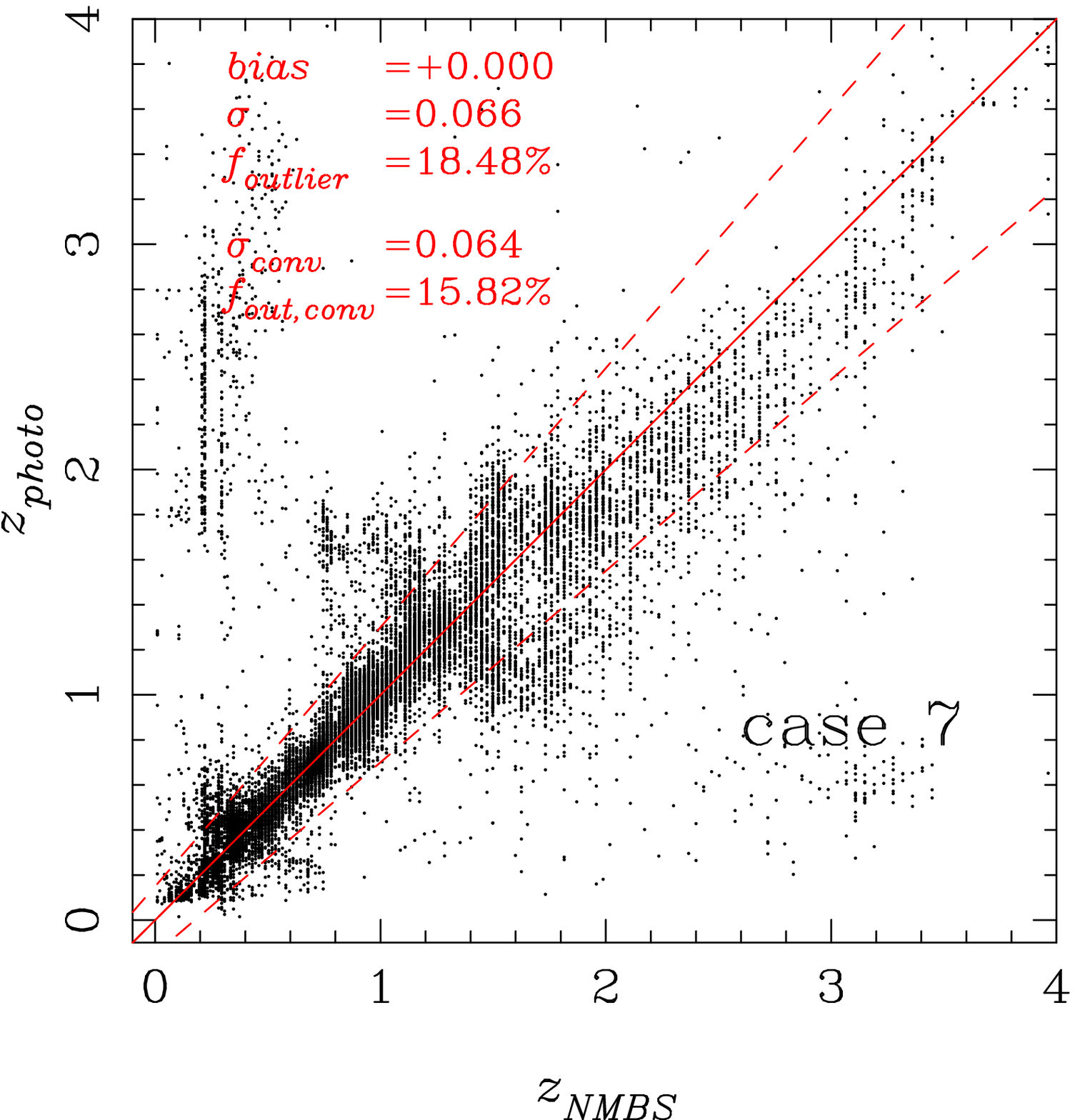}
\caption{
  Photometric redshifts based on the $griz$ photometry for the $K$-selected NMBS catalog.
}
\label{fig:nmbs_photoz}
\end{figure*}
%---------------------

The physical priors and template error functions introduced in the paper
do not in principle depend on
which filter is used for object detections, but it is still important to
explicitly show that they work for a catalog that is not $i$-selected.
For this, we use the NMBS catalog used in Section 8, in which the objects are
detected in the $K$-band.
Fig. \ref{fig:nmbs_photoz} shows photo-$z$'s for case 1, 2, and 7 using
the $griz$ photometry.
In case 7, only the physical priors and template error functions 
with offsets are applied (i.e., no $N(z)$ prior).
We do not show the cases that involve the $N(z)$ prior because
the $N(z)$ prior is for the $i$-band and thus it by construction does not work
well for a $K$-selected sample.  The statistics are summarized in Table \ref{tab:photoz_ksel}.

A comparison between case 1 and 2 clearly demonstrate that the priors work 
very well for the $K$-selected catalog; the up-scattered population at low-$z$
is strongly suppressed and the overall scatter is reduced.  In addition,
the template error functions applied in case 7 further improve the accuracy.
In particular, the bias is significantly reduced and there is  no remaining
systematic offset.  Based on these results, we conclude that the physical priors
and template error functions are effective regardless of the detection filter.

%---------------------------------
\begin{table*}[htb]
  \begin{center}
    \begin{tabular}{cccc|ccccc}
      case & physical priors & $N(z)$ prior & template errfn   & bias     & $\sigma$& $f_{outlier}$ & $\sigma_{conv}$ & $f_{outlier,conv}$\\\hline
      1    & No              & No           & No               & $-0.025$ & $0.090$ & $27.2\%$    & $0.107$       & $31.3\%$\\
      2    & Yes             & No           & No               & $-0.032$ & $0.063$ & $20.5\%$    & $0.076$       & $18.1\%$\\
      7    & Yes             & No           & Yes (w/  offset) & $+0.000$ & $0.066$ & $18.5\%$    & $0.064$       & $15.8\%$\\
      \hline
    \end{tabular}
  \end{center}
  \caption{
    Summary of photo-$z$ accuracies for the $K$-selected NMBS catalog using $griz$ photometry.
  }
  \label{tab:photoz_ksel}
\end{table*}
%---------------------------------

%-------------------------------------------------------------------
\section{Effects of applying 'wrong' physical priors}

%---------------------
\begin{figure*}
\epsscale{1}
\plottwo{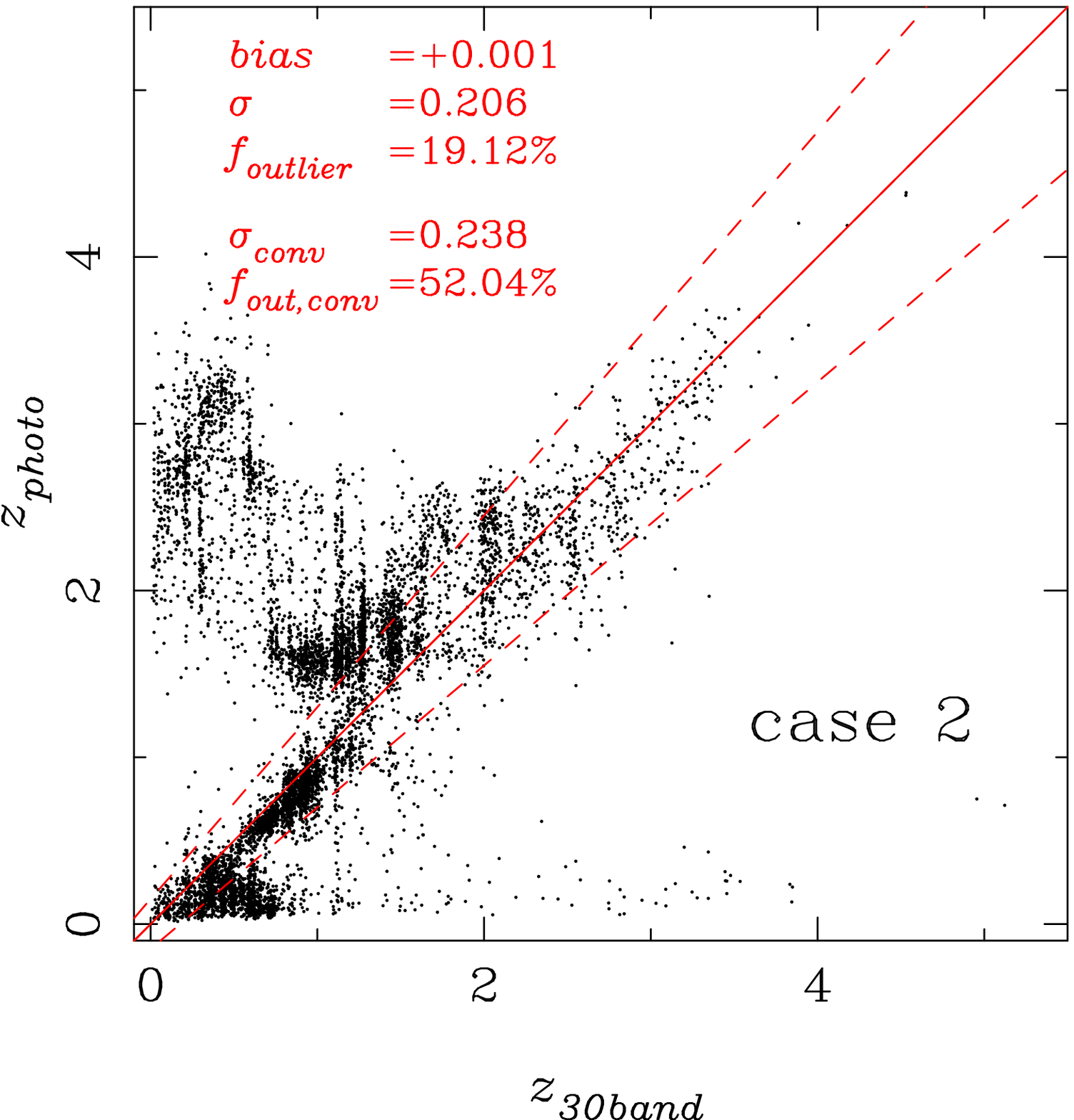}{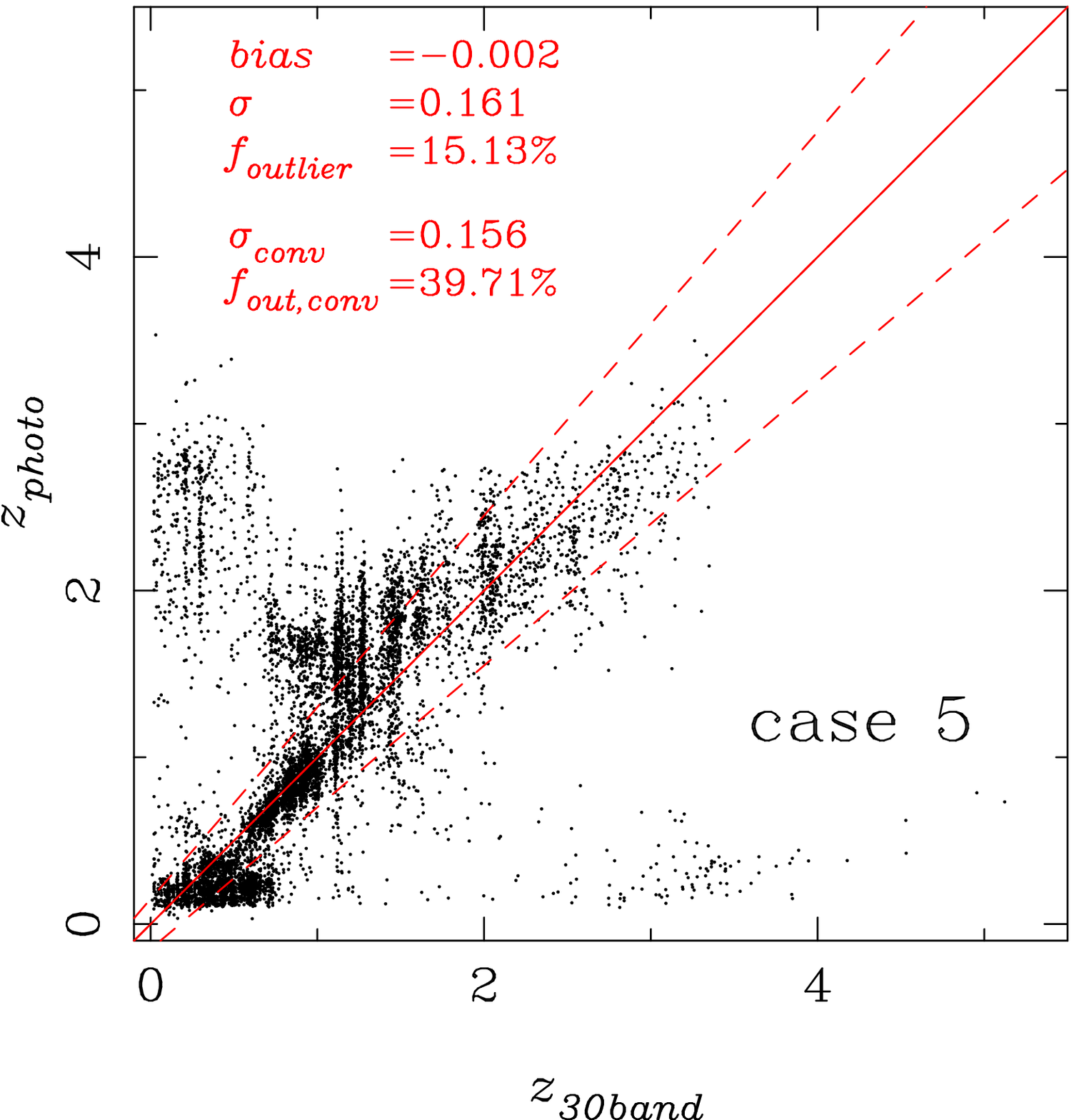}\vspace{0.2cm}
\caption{
  Photometric redshifts computed using the 'wrong' physical priors for case 2 and 5.
}
\label{fig:wrong_priors}
\end{figure*}
%---------------------

Our physical priors are motivated by observations, but it would be instructive
to show what happens if we change the prior parameters.
As an illustrative example, we change $SFR^*(z)$ defined in Eq. \ref{photoz:eq:sfr_star2} by
$-1$ dex.  What it effectively does is to shift the observed sequence of
star forming galaxies downwards by $\Delta SFR=-1$~dex in the SFR vs stellar mass
prior defined in Section 4.3.  It also changes the attenuation prior in Section
4.4 in the sense that the prior has larger attenuation than what is observed.

We run our code with these 'wrong' priors on the same COSMOS catalog as in Section 6.
Here, we keep the $N(z)$ prior and template error functions unchanged.
The resultant photo-$z$'s for case 2 and 5 are shown in Fig.
\ref{fig:wrong_priors}.  Statistics for all the cases are summarized in Table \ref{tab:photoz_wrong_priors}.
Case 1 and 6 do not use the physical priors, but their accuracies are reproduced
from Table \ref{tab:photoz} for easy comparisons.
It is clear that the wrong priors degrade photo-$z$'s.  The priors give underestimated
photo-$z$'s at $z\lesssim0.5$ and overestimated ones at $z\sim1$.
This degradation is expected because the priors give a large probability to objects
that do not really exist in the universe.  In fact, by comparing with case 5 and 6
in Table \ref{tab:photoz_wrong_priors}, it is better not to apply the wrong physical priors.
The improvement delivered by our fiducial physical priors shown in Section 6.3 in turn suggests that
thes fiducial priors are not far from optimal.

%---------------------------------
\begin{table*}[htb]
  \begin{center}
    \begin{tabular}{cccc|ccccc}
      case & physical priors & $N(z)$ prior & template errfn   & bias     & $\sigma$& $f_{outlier}$ & $\sigma_{conv}$ & $f_{outlier,conv}$\\\hline
      1    & No              & No           & No               & $+0.011$ & $0.143$ & $30.9\%$    & $0.182$       & $45.2\%$\\
      2    & Yes             & No           & No               & $+0.001$ & $0.206$ & $19.1\%$    & $0.238$       & $52.0\%$\\
      3    & Yes             & Yes          & No               & $-0.025$ & $0.166$ & $15.6\%$    & $0.173$       & $41.8\%$\\
      4    & Yes             & Yes          & Yes (w/o offset) & $-0.017$ & $0.163$ & $14.9\%$    & $0.163$       & $40.2\%$\\
      5    & Yes             & Yes          & Yes (w/  offset) & $-0.002$ & $0.161$ & $15.1\%$    & $0.156$       & $39.7\%$\\
      6    & No              & Yes          & Yes (w/  offset) & $+0.004$ & $0.101$ & $19.1\%$    & $0.095$       & $25.8\%$\\
      \hline
    \end{tabular}
  \end{center}
  \caption{
    Summary of photo-$z$ accuracies with wrong priors.
  }
  \label{tab:photoz_wrong_priors}
\end{table*}
%---------------------------------

\rm

%-------------------------------------------------------------------

\end{document}